\documentclass[aps,prl,twocolumn,superscriptaddress]{revtex4-2}
\usepackage{bm}
\usepackage{amsmath,amssymb,amsfonts}%
\usepackage{siunitx}
\usepackage{chemarrow}
\usepackage{nicefrac}
\usepackage{braket}
\usepackage{graphicx}
\usepackage{xcolor}
\usepackage{times}

\usepackage{ulem}

\newcommand{\jzdel}[1]{{\color{red}\ifmmode\text{\sout{\ensuremath{#1}}}\else\sout{#1}\fi}}

\newcommand{\ccdel}[1]{{\color{red}\ifmmode\text{\sout{\ensuremath{#1}}}\else\sout{#1}\fi}}

\newcommand{\lhdel}[1]{{\color{red}\ifmmode\text{\sout{\ensuremath{#1}}}\else\sout{#1}\fi}}

\usepackage[unicode=true,pdfusetitle, bookmarks=true,bookmarksnumbered=false,bookmarksopen=false,breaklinks=false,pdfborder={0 0 0},backref=false,colorlinks=true,linkcolor=blue,citecolor=blue,urlcolor=blue]{hyperref}

\begin{document}
\title{Two-particle quantum interference in a nonlinear optical medium:\\ a witness of timelike indistinguishability}

\author{Chao Chen}
\altaffiliation{These authors contributed equally to this work.}
\affiliation{National Laboratory of Solid State Microstructures and School of Physics, Nanjing University, Nanjing 210093, China}
\affiliation{Collaborative Innovation Center of Advanced Microstructures, Nanjing University, Nanjing 210093, China}
\affiliation{School of Physical Science and Technology, Ningbo University, Ningbo 315211, China}
\author{Shu-Tian Xue}
\altaffiliation{These authors contributed equally to this work.}
\affiliation{National Laboratory of Solid State Microstructures and School of Physics, Nanjing University, Nanjing 210093, China}
\affiliation{Collaborative Innovation Center of Advanced Microstructures, Nanjing University, Nanjing 210093, China}
\author{Yu-Peng Shi}
\affiliation{National Laboratory of Solid State Microstructures and School of Physics, Nanjing University, Nanjing 210093, China}
\affiliation{Collaborative Innovation Center of Advanced Microstructures, Nanjing University, Nanjing 210093, China}
\author{Jing Wang}
\affiliation{National Laboratory of Solid State Microstructures and School of Physics, Nanjing University, Nanjing 210093, China}
\affiliation{Collaborative Innovation Center of Advanced Microstructures, Nanjing University, Nanjing 210093, China}
\author{Zi-Mo Cheng}
\affiliation{National Laboratory of Solid State Microstructures and School of Physics, Nanjing University, Nanjing 210093, China}
\affiliation{Collaborative Innovation Center of Advanced Microstructures, Nanjing University, Nanjing 210093, China}
\author{Pei Wan}
\affiliation{National Laboratory of Solid State Microstructures and School of Physics, Nanjing University, Nanjing 210093, China}
\affiliation{Collaborative Innovation Center of Advanced Microstructures, Nanjing University, Nanjing 210093, China}
\author{Zhi-Cheng Ren}
\affiliation{National Laboratory of Solid State Microstructures and School of Physics, Nanjing University, Nanjing 210093, China}
\affiliation{Collaborative Innovation Center of Advanced Microstructures, Nanjing University, Nanjing 210093, China}
\author{Michael G. Jabbour}
\email{mjabbour@telecom-sudparis.eu}\affiliation{SAMOVAR, T\'el\'ecom SudParis, Institut Polytechnique de Paris, 91120 Palaiseau, France}
\affiliation{Centre for Quantum Information and Communication, Ecole polytechnique de Bruxelles, Universit\'{e} libre de Bruxelles, 1050 Bruxelles, Belgium}
\affiliation{Department of Physics, Technical University of Denmark, 2800 Kongens Lyngby, Denmark}
\author{Nicolas J. Cerf}
\email[]{nicolas.cerf@ulb.be}
\affiliation{Centre for Quantum Information and Communication, Ecole polytechnique de Bruxelles, Universit\'{e} libre de Bruxelles, 1050 Bruxelles, Belgium}
\author{Xi-Lin Wang}
\email[]{xilinwang@nju.edu.cn}
\affiliation{National Laboratory of Solid State Microstructures and School of Physics, Nanjing University, Nanjing 210093, China}
\affiliation{Collaborative Innovation Center of Advanced Microstructures, Nanjing University, Nanjing 210093, China}
\affiliation{Hefei National Laboratory, Hefei 230088, China}
\affiliation{Synergetic Innovation Center of Quantum Information and Quantum Physics, University of Science and Technology of China, Hefei 230026, China}
\author{Hui-Tian Wang }
\email[]{htwang@nju.edu.cn}
\affiliation{National Laboratory of Solid State Microstructures and School of Physics, Nanjing University, Nanjing 210093, China}
\affiliation{Collaborative Innovation Center of Advanced Microstructures, Nanjing University, Nanjing 210093, China}
\affiliation{Collaborative Innovation Center of Extreme Optics, Shanxi University, Taiyuan 030006, China}

\date{\today}

\begin{abstract}
\textbf{The Hong-Ou-Mandel effect is a paradigmatic quantum phenomenon demonstrating the interference of two indistinguishable photons that are linearly coupled at a 50:50 beam splitter. Here, we transpose such a two-particle quantum interference effect to the nonlinear regime, when two single photons are impinging on a parametric down-conversion crystal. Formally, this transposition amounts to exchanging space and time variables, giving rise to an unknown  form of timelike quantum interference.
The two-photon component of the output state is a superposition of the incident photons being either transmitted or reborn, that is, replaced by indistinguishable substitutes due to their interaction with the nonlinear crystal. We experimentally demonstrate the suppression of the probability of detecting precisely one photon pair when the amplification gain is tuned to 2, which arises from the destructive interference between the transmitted and reborn photon pairs. This heretofore unobserved quantum manifestation of indistinguishability in time pushes nonlinear quantum interference towards a new regime with multiple photons. Hence, composing this effect with larger linear optical circuits should provide a tool to generate multimode quantum non-Gaussian  states, which are essential resources for photonic quantum computers.}  
\end{abstract}





\maketitle
\begin{figure}[t!]
\includegraphics[width=1\columnwidth]{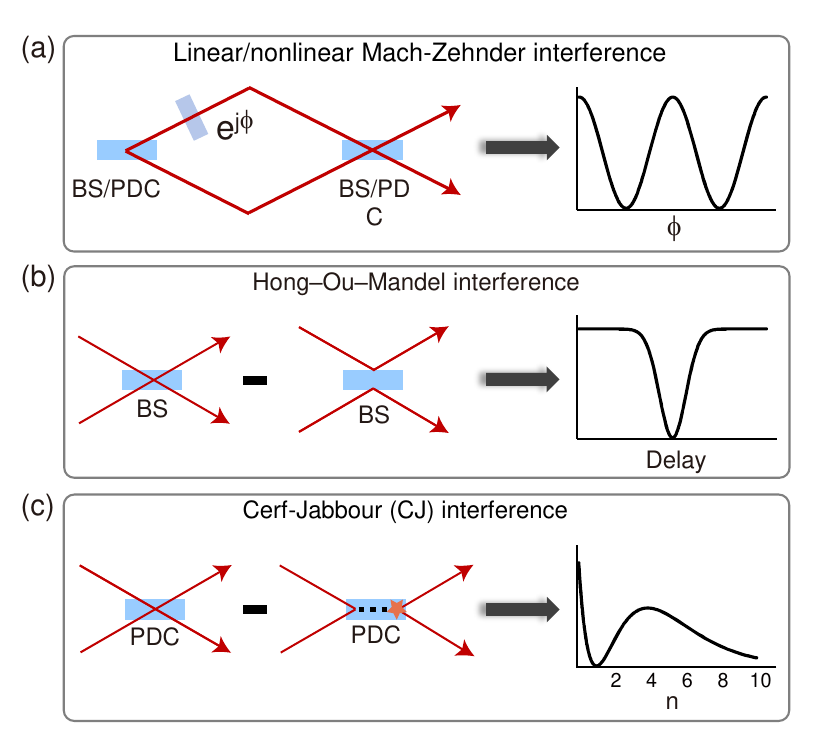}
\caption{Quantum interferometer configurations. (a),~Mach-Zehnder (MZ) interferometer: following the first beam splitter (BS), photons travel through two possible optical paths and interfere at another BS. The light intensity at any output oscillates with the relative phase between the paths. If the BS are replaced with parametric down-conversion (PDC) crystals, we obtain a SU(1,1) nonlinear interferometer and phase-sensitive stimulated emission is observed. (b),~Hong-Ou-Mandel (HOM) interference: two single photons impinge on a BS from different input modes. The photons being both transmitted interfere destructively with those being both reflected, leading to vanishing coincidence at the output for a splitting ratio~1/2. (c),~Cerf-Jabbour (CJ) interference: two incident single photons interfere at a PDC crystal. The photons evolve into a superposition of being either directly transmitted or being reborn in the PDC (replaced by indistinguishable substitutes via up- and down-conversion). The interference between these two components cancels the probability of outputting precisely one pair of photons for a gain 2.}
\label{fig:1}
\end{figure}

Quantum interference lies at the heart of fundamental quantum mechanics \cite{mandelQuantumEffectsOnephoton1999}. One typical optical interference configuration is the Mach-Zehnder (MZ) interferometer (Fig.~\ref{fig:1}a), which shows oscillatory interference fringes due to the relative phase between the upper and lower optical paths. The linear MZ interferometer can be extended to SU(1,1) nonlinear quantum interferometry \cite{yurkeSUSUInterferometers1986b} by replacing the beam splitters~(BSs) with parametric down-conversion (PDC) crystals. PDC-based MZ interferometers shape the fields of induced coherence by path identity \cite{wangInducedCoherenceInduced1991,zouInducedCoherenceIndistinguishability1991,heuerInducedCoherenceVacuum2015a,onoObservationNonlinearInterference2019,hochrainerQuantumIndistinguishabilityPath2022,fengOnchipQuantumInterference2023} or stimulated emission \cite{simonOptimalQuantumCloning2000,lamas-linaresExperimentalQuantumCloning2002,sunStimulatedEmissionResult2007a}, and have shown attractive potential application in quantum imaging \cite{lemosQuantumImagingUndetected2014,qianQuantumInducedCoherence2023} and quantum metrology \cite{qinUnconditionalRobustQuantum2023} by exploiting the phase-sensitive output. In comparison, the celebrated Hong-Ou-Mandel (HOM) effect \cite{hongMeasurementSubpicosecondTime1987a} relies on phase-independent quantum interference when two indistinguishable photons interfere at a 50:50~BS (Fig.~\ref{fig:1}b). The coincidence of detecting one photon at each output port vanishes because two photons being both transmitted or both reflected interfere destructively. Being of major fundamental interest, the HOM effect has been demonstrated not only with photons \cite{santoriIndistinguishablePhotonsSinglephoton2002,patelTwophotonInterferenceEmission2010a, kobayashiFrequencydomainHongOu2016a} but also with surface plasmons \cite{fakonasTwoplasmonQuantumInterference2014}, atoms \cite{lopesAtomicHongOu2015} and phonons \cite{toyodaHongOuMandel2015}. Furthermore, the HOM effect has aroused the interference of multiple quantum light sources \cite{wangExperimentalTenPhotonEntanglement2016,menssenDistinguishabilityManyParticleInterference2017a}, which have applications in quantum computation \cite{broomePhotonicBosonSampling2013c,tillmannExperimentalBosonSampling2013b,crespiIntegratedMultimodeInterferometers2013b,wangHighefficiencyMultiphotonBoson2017,zhong12PhotonEntanglementScalable2018,zhongQuantumComputationalAdvantage2020}.

Intriguingly, when transposing the HOM effect to the nonlinear regime, an overlooked quantum interference mechanism emerges \cite{cerfTwobosonQuantumInterference2020}, which can in principle be observed when two single photons impinge on a PDC crystal (Fig.~\ref{fig:1}c). 
Owing to the creation and annihilation terms in the PDC Hamiltonian~\eqref{eq:updc}, the output photons can be the two originally incident ones or, instead, two newly generated photons (meanwhile the incident ones are annihilated), which we call reborn photons. Remember that the HOM effect stems from the Bose statistics of indistinguishable photons. Likewise here, the transmitted photon pair destructively interferes with the reborn photon pair, leading to a depletion of the probability of outputting one photon pair. This two-photon nonlinear interference in a PDC, which we dub Cerf-Jabbour (CJ) interference \cite{cerfTwobosonQuantumInterference2020}, is a striking manifestation of indistinguishability in time (roughly speaking, one cannot recognize whether the output photons are ``the same as'' the input photons or not).

Although multi-photon nonlinear interferences have been reported in the literature (see, e.g., \cite{chekhovaNonlinearInterferometersQuantum2016,onoObservationNonlinearInterference2019,fengOnchipQuantumInterference2023}), such experiments belong to the class depicted in Fig.~\ref{fig:1}(a), which exploits path-identity
induced coherence \cite {hochrainerQuantumIndistinguishabilityPath2022}. 
Here, we demonstrate the two-photon CJ interference, which fundamentally differs from previous experiments as it is the nonlinear counterpart to the HOM experiment. To match a 50:50 BS, the gain $g$ of the PDC should be tuned to $2$, such that up- and downconversion are simultaneously at play on a fifty-fifty basis. We achieve this strong interaction of photons (very high $g$) via a PDC process pumped by an ultra-tightly focused high-power femtosecond laser that has a meticulous mode matching with heralded single photons in all degrees of freedom. Moreover, we successfully retrieve photon-number distribution through six-channel coincidence measurement and post-processing via a direct inverse technique and a fitted experimental model, which enables us to observe a depletion of the probability of outputting precisely one pair of photons and demonstrate nonlinear destructive interference.







\begin{figure*}
\includegraphics[width=1.9\columnwidth]{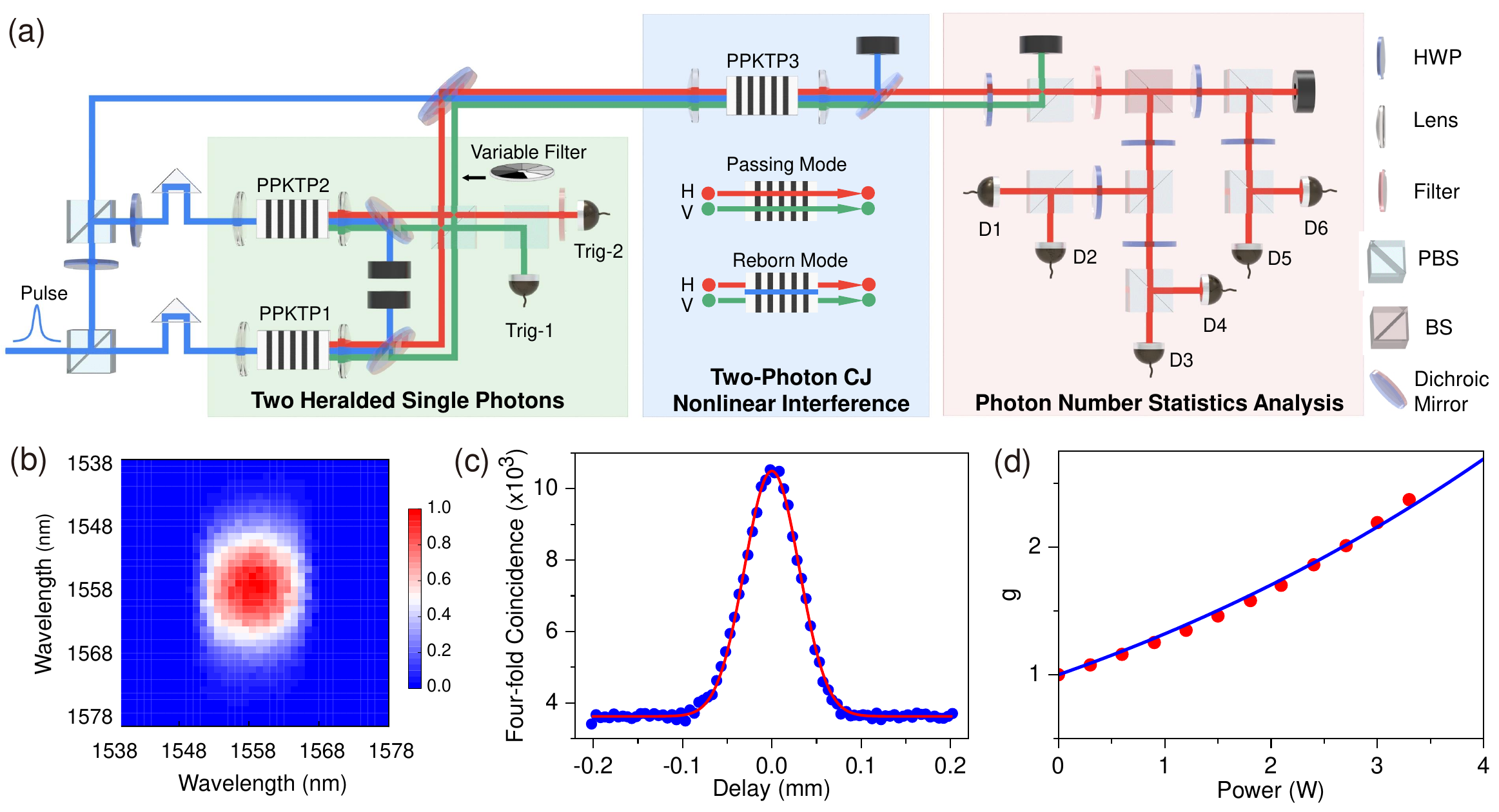}
\caption{Experimental setup and PDC source. (a), A 779 nm pulsed laser with pulse width of \qty{140}{fs} and repetition frequency of 80MHz is divided by beam splitters to pump three PPKTPs with length of \qty{2.5}{mm}, which support type-II parametric down conversion from a 779 nm horizontally polarized photon to a pair of collinear photons with horizontal (H) and vertical (V) polarization, respectively pictured in red and green. The heralded H- and V-polarized single photons generated from PPKTP1 and PPKTP2 are spatially and temporally matched with the PDC modes of PPKTP3, where interference occurs. The photon number distribution of the H-polarized output mode of PPKTP3 is analyzed by an array of superconducting nanowire single-photon detectors. (b), Measured joint spectrum when a 15 nm filter is introduced for the H-polarized photons. (c), Temporal match of the heralded H-polarized single photons with the PDC photons in PPKTP3, verified by observing stimulated emission. When the heralded mode is temporally overlapped with the PDC mode in PPKTP3, the coincidence rate of detector 1-3 and detector Trig-1 is amplified. The time delay is adjusted by using the prisms before the PPKTP1 and PPTKP2. Coincidence rate including detector Trig-2 is measured when calibrating the temporal match of the heralded $V$-polarized single photon. (d), Measured parametric gain $g$ 
(red dots) of PPKTP3 when varying the pump power. The dashed line shows the theoretical values of $g$ assuming that $r$ is proportional to the square root of pump power \cite{lvovskySqueezedLight2016}.}
\label{fig:setup}
\end{figure*}

Our experiment thus hints that timelike indistinguishability has very unsuspected consequences and pushes nonlinear quantum interference \cite{yurkeSUSUInterferometers1986b,paterovaNonlinearInterferenceCrystal2020,luoQuantumOpticalCoherence2021} to a multi-photon regime, necessitating nonlinear interaction of single photons. Aside from the fundamental physics implications, we also show that a strongly non-Gaussian Wigner-negative regime could be accessed by integrating this CJ nonlinear interference within a large-scale linear optical interferometer based on a cascade network, which would be valuable for the development of photonic quantum computers \cite{mariPositiveWignerFunctions2012, veitchEfficientSimulationScheme2013,rahimi-keshariSufficientConditionsEfficient2016}. 



\paragraph{\bf Experiment.}
In our experiment, shown in Fig.~\ref{fig:setup}, the core unit -- the parametric amplifier -- is a type-II periodically poled potassium titanyl phosphate (PPKTP) crystal, which allows a horizontally polarized pump photon at 779 nm to be down-converted into a pair of photons at 1558 nm being horizontally and vertically polarized respectively. The unitary operation of the PPKTP reads
\begin{equation}
    U_g^{\text{PDC}}=\text{exp}[r(\hat{a}_\text{H}\hat{a}_\text{V}-\hat{a}_\text{H}^\dagger\hat{a}_\text{V}^\dagger)],
    \label{eq:updc}
\end{equation}
where $\hat{a}_\text{H(V)}$ is the signal (idler) mode operator with horizontal (vertical) polarization and $r$ is the squeezing parameter, related to the parametric gain $g$ via $g=\cosh^2{r}$. Photons are created (annihilated) by pairs as a result of down-conversion (up-conversion) of pump photons due to the structure of $U_g^{\text{PDC}}$, and we denote as $P_n$ the probability of observing $n$ pairs at the output (\textit{i.e.}, the output state $\ket{n,n}$). The gain $g$ governs the depth of the two-photon CJ nonlinear interference, just like the splitting ratio of a BS determines the visibility of HOM interference. When $g=2$, two-photon events vanish due to fully destructive interference, that is, we should observe $P_1=0$ in ideal conditions (see Supplementary Information).

In the experimental setup shown in Fig.~\ref{fig:setup}(a), a femtosecond pulsed laser beam (blue) is divided by beam splitters to pump three identical PPKTP crystals coherently. The heralded two single photons are generated through spontaneous PDC processes in PPKTP1 and PPKTP2 at low pump power of \qty{100}{mW}. To match the polarization states of the PDC photons in PPKTP3, heralded H- (in red) and V-polarized (in green) single photons are generated in PPKTP1 and PPKTP2, respectively, by triggering their counterpart photons. We combine the two heralded H- and V-polarized single photons at a polarization beam splitter (PBS). Then, the two single photons are aligned to interact with PPKTP3, whose output is analyzed by multi-channel single-photon coincidence measurement. To achieve high gain for PPKTP3, the laser with tunable power up to~$\qty{3}{W}$ is tightly focused on the crystal. The beam waist of the pump laser is~$\qty{30}{\mu m}$, which is smaller than that was employed in \textit{Jiuzhang 2.0} quantum computer\cite{zhongPhaseProgrammableGaussianBoson2021a}.

We optimize our experimental setup to realize mode match between the two heralded single photons and the PDC photon pair in PPKTP3 in spectral, spatial and temporal degrees of freedom. The spectral indistinguishability is enforced in two ways: (1) all three PPKTP crystals are temperature controlled to generate down-converted photon pairs with degenerate central wavelength at 1558 nm; (2) the side lobes in joint spectrum are removed by a 15-nm filter, so that the two modes are frequency uncorrelated, as shown in Fig.~\ref{fig:setup}(b). The average pairwise purity estimated by unheralded second order correlation is 0.92 (\cite{christProbingMultimodeSqueezing2011,zhongQuantumComputationalAdvantage2020} and see Supplementary Information). The spatial modes of the heralded photons are carefully aligned with those of the down-converted photons from PPKTP3. As for temporal match, the arriving time of the heralded single photons at PPKTP3 is adjusted by two delay lines (two prisms on motorized stages). For example, when the H-polarized single photon is temporally overlapped with the H-polarized PDC mode in PPKTP3, we verify that the emission of down-converted photons is stimulated, as shown in Fig.~\ref{fig:setup}(c). To clearly confirm temporal mode matching, both PPKTP1/2 and PPKTP3 are pumped by laser beams of high power for obtaining a remarkable stimulated emission peak. With all the above optimizations, the overall mode match of the heralded single photon and the PDC modes in PPKTP3 is about 0.65 (see Supplementary Information), which outperforms the result in \cite{qianMultiphotonNonlocalQuantum2023} and is sufficient to clearly show the two-photon suppression due to the CJ nonlinear interference.

To observe the suppression of the probability $P_1$ of outputting one pair, which is a direct evidence of the two-photon CJ nonlinear quantum interference, analysis of photon number distribution is of the essence. Provided the heralded photons impinge PPKTP3 in pairs, the output photon number distributions of the H- and V-polarized modes are in principle symmetric, rendering measurement of only one of these two modes sufficient to analyze the photon-pair probability distribution. For resource saving, we uniformly distribute the H-polarized output photons into six superconducting nanowire single-photon detectors with average detection efficiency of $80\%$, and leave the V-polarized photons undetected. By analyzing the measured multi-channel coincidence, we use direct inverse method (see Supplementary Information) to obtain the probability of outputting one H-polarized photon, which we associate with $P_1$. 
When the pump laser of PPKTP1 and PPKTP2 is blocked, the output state of PPKTP3 is a two-mode squeezed vacuum state. In Fig.~\ref{fig:setup}(d), we show the parametric gain $g$ as deduced from our multi-channel coincidence measurement, plotted against the pump power of PPKTP3.

\begin{figure*}
\includegraphics[width=1.9\columnwidth]{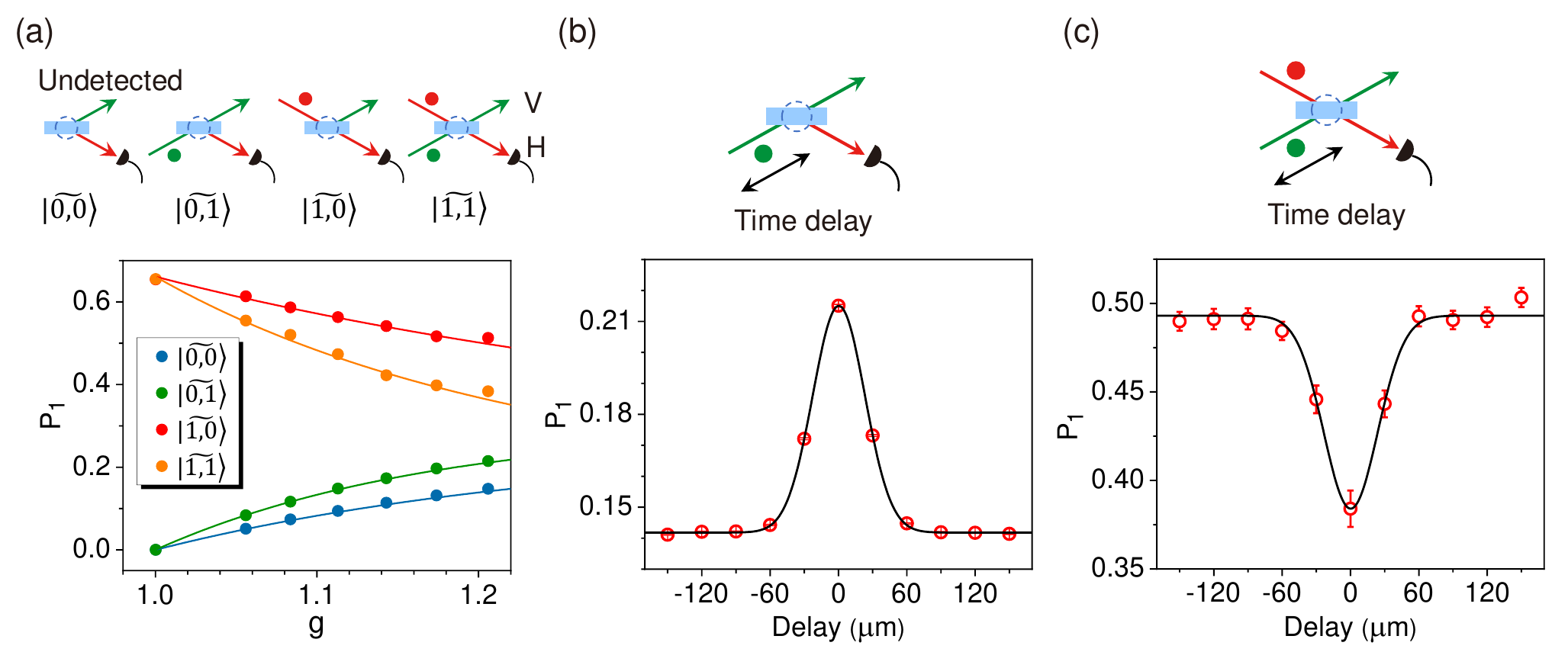}
\caption{Measured probability $P_1$ of outputting one horizontally polarized photon. (a), Measured $P_1$ with parametric gain ranging from $g=1$ to $g\approx 1.2$ for four input states, corresponding to either of the heralded single photons being blocked or not. (b), Measured $P_1$ when the time delay of the $V$-polarized single photon varies, while the $H$-polarized single photon is blocked. (c), Measured $P_1$ when the time delay of the $V$-polarized single photon varies, while the $H$-polarized single photon is temporally matched. The parametric gains for (b) and (c) are fixed at $g\approx 1.2$.
}
\label{fig:3}
\end{figure*}

\paragraph{\bf Results.}
For different input states, the measured probability of outputting one H-polarized photon $P_1$ is shown in Fig.~\ref{fig:3}(a). Depending on whether the heralded H- or V-polarized single photon are blocked or not, we label the input state as $\ket{\widetilde{0,0}}$,  $\ket{\widetilde{0,1}}$, $\ket{\widetilde{1,0}}$, and $\ket{\widetilde{1,1}}$. The tilde are used to stress that, due to imperfect mode match, these states could deviate from the ideal input states denoted as $\ket{0,0}$, $\ket{0,1}$, $\ket{1,0}$, and $\ket{1,1}$. The parametric gain $g$ ranges from 1 to 1.21, corresponding to increase the pump power from 0 to \qty{700}{mW}. When no single photons are input ($\ket{\widetilde{0,0}}$), it is common to see that the larger pump power gives rise to higher generation rate of down-converted photons, \textit{i.e.}, $P_1$ increases with $g$. As expected,  $P_1$ increases even faster when  inputting a V-polarized single photon ($\ket{\widetilde{0,1}}$) as a result of stimulated emission. In contrast, if inputting a H-polarized photon ($\ket{\widetilde{1,0}}$), $P_1$ decreases with $g$ because the probability of outputting more than one photons emerges in the active PDC process. Note that for $g=1$ (when the pump laser of PPKTP3 is blocked and no interaction occurs), we have $P_1\approx 0.65$, deviating from the theoretical value $P_1=1$ as expected for ideal single-photon input state $\ket{1,0}$. This is mainly due to the limited mode match between the heralded single photon and the PDC modes in PPKTP3, which leads to mix some state $\ket{0,0}$ together with states $\ket{1,0}$ in the initial state $\ket{\widetilde{1,0}}$ (see Supplementary Information). 

Now, the key observation is that $P_1$ decays even further with $g$ when both H- and V-polarized heralded single photons are inputted ($\ket{\widetilde{1,1}}$). It is indeed counterintuitive that inputting an extra V-polarized photon results in the decline (instead of the enhancement) of the probability of outputting a single H-polarized photon. This indicates that a two-photon destructive quantum interference is at play when single photons are injected in pair. In Fig.~\ref{fig:3}(b) and (c), we compare the measured $P_1$ for single-photon stimulation and photon pair suppression (corresponding, respectively, to inputs $\ket{\widetilde{0,1}}$ and $\ket{\widetilde{1,1}}$) when the injected photons arrive at PPKTP3 together with the pump pulse. When the V-polarized single photon is gradually temporally mismatched by changing the time delay, the input state deviates to $\ket{\widetilde{0,0}}$ ($\ket{\widetilde{1,0}}$) so that the stimulated peak (suppressed dip) subsides, as shown in Fig.~\ref{fig:3} (b) (Fig.~\ref{fig:3}(c)).

\begin{figure*}
\includegraphics[width=1.9\columnwidth]{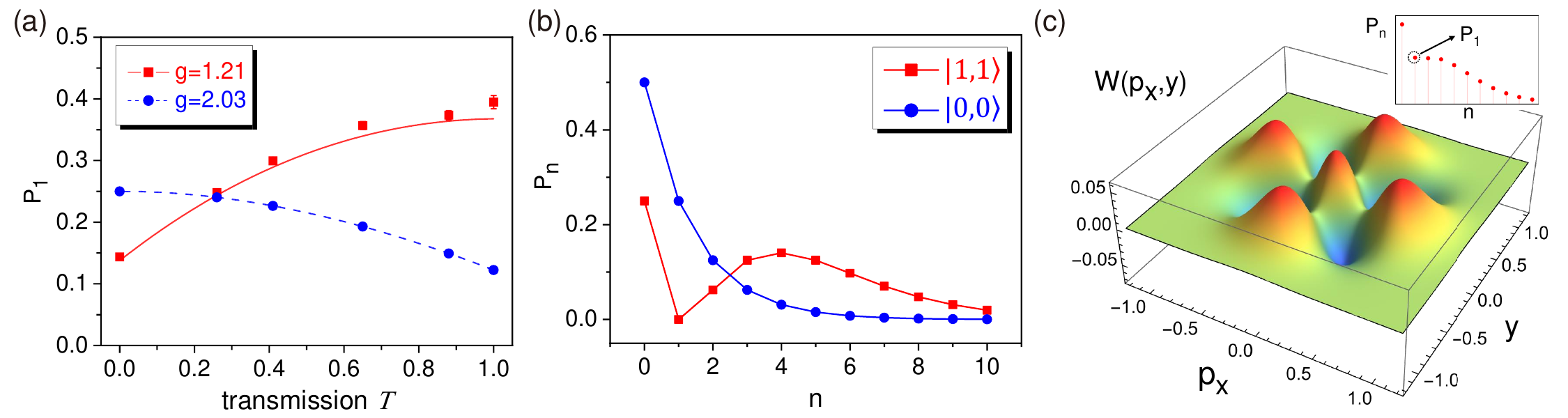}
\caption{ Characteristics of the output state resulting from two-photon CJ nonlinear interference.  (a),~Measured $P_1$ versus the transmission $\mathrm{\it{T}}$ of a variable neutral filter inserted before PPKTP3 to tune the heralding efficiency of the incident single-photon pair. When $g=1.21$, the deduced $P_1$ (red line) from the fitted model agrees well with the approximated $P_1$ (red squares) from multi-channel coincidence measurement, confirming the validity of our fitted model. In this case, $P_1$ ramps up as $\mathrm{\it{T}}$ increases. In contrast, at high gain of $g=2.03$, the deduced $P_1$ is progressively suppressed when the transmission $\mathrm{\it{T}}$ tends to 1. 
(b),~Theoretically simulated output photon-number distribution $P_n$ with either vacuum or perfect single photon pair input states when $g=2$. For $\ket{1,1}$ input state, $P_1$ vanishes due to the two-photon CJ nonlinear interference in the PDC crystal. (c),~Experimentally reconstructed Wigner function $W (p_x,y)$ with $x=0$ and $p_y=0$ of the output two-mode state when $U_{g=2.03}^{\text{PDC}}$ is applied on the input $\ket{\widetilde{1,1}}$ state. The corresponding photon-number distribution retrieved from the experiment confirms a depletion of $P_1$, as shown in the inset.
}
\label{fig:4}
\end{figure*}

\paragraph{\bf Discussion.}
We now focus on the two-photon CJ interference in a PDC crystal and analyze its main features in Fig.~\ref{fig:4}. First, the impact of the quality of the injected photon pair is analyzed and compared at different values of the parametric gain $g$, which is the analogue of the split ratio of the BS affecting the visibility of the HOM dip, as shown in Fig.~\ref{fig:4}(a). A neutral density filter with step-variable transmission $\mathrm{\it{T}}$ is introduced in the front of PPKTP3 to adjust the heralding efficiency of the input single photon pairs. For $\mathrm{\it{T}}=1$, the input state mostly consists of $\ket{1,1}$; for $\mathrm{\it{T}}=0$, the input state is a vacuum state although the detectors Trig-1 and Trig-2 count single photons. At small gain ($g=1.21$), $P_1$ consistently increases with $\mathrm{\it{T}}$ as more single photon pairs are inputted, because the directly transmitted photon pairs dominate those interacting with the PDC crystal. Amazingly, this explains why the two-photon CJ nonlinear quantum interference had remained unsuspected in all previous experiments with moderate pump power. Here, we promote the gain to $g=2.03$  
and observe that inputting more photon pairs instead comes with outputting less photon pairs, demonstrating the destructive nonlinear quantum interference in the PPKTP crystal. In practice, the residual value of $P_1 \approx 0.1$ at $\mathrm{\it{T}}=1$ mainly results from the limited mode matching, as mentioned before.

Importantly, together with the interferometric suppression of $P_1$, the entire photon-number distribution of the output state is impacted by the CJ effect \cite{cerfTwobosonQuantumInterference2020}. In Fig.~\ref{fig:4}(b), we show the theoretically simulated photon-number distribution $P_n$ of the $H$-polarized mode. For the $\ket{0,0}$ input, the output is a two-mode squeezed vacuum state, exhibiting an exponential decay of $P_n$ with $n$ as expected. With ideal one-photon-pair input state $\ket{1,1}$, the output photon number distribution shows a dip at $n=1$ due to the two-photon CJ nonlinear quantum interference. In Fig.~\ref{fig:4}(c), we also plot the reconstructed output two-mode state in phase space, together with the corresponding photon-number distribution retrieved from the experiment. Although experimental imperfections lead to  some leftover probability $P_1$, the output state still remains quite non-Gaussian, with the presence of Wigner negativity. The non-Gaussianity of the output state stems from the input single photons, in contrast with the reported photon addition/subtraction protocols \cite{wengerNonGaussianStatisticsIndividual2004,jeongGenerationHybridEntanglement2014,raNonGaussianQuantumStates2020a,walschaersNonGaussianQuantumStates2021}. When increasing the pump power to $g=3$, four-photon destructive interference in the PDC crystal results in the suppression of $P_2$ (see Supplementary Information).

\paragraph{\bf Conclusion.}

We experimentally demonstrate a surprising quantum interference mechanism by impinging two single photons on a highly-pumped PDC crystal, which is the nonlinear counterpart to the well-known HOM interference when timelike indistinguishability is at play. We observe a depletion of the probability of detecting precisely one pair of photons at the output, witnessing the destructive interference between the transmitted photons and their reborn substitutes due to the nonlinear interaction when the gain reaches 2. Aside from the fundamentally new mechanism that it unveils,
our work unambiguously verifies the two-photon rebirth (a combination of generating and annihilating photons in pairs) and pushes nonlinear quantum interference to genuine multi-photon regime, which could enrich the toolkit for nonlinear interaction between single photons and matter~\cite{hubelDirectGenerationPhoton2010,guerreiroNonlinearInteractionSingle2014}. The resulting ability to control the photon-number distribution of the output two-mode state provides an alternative
powerful scheme to engineer non-Gaussian entangled states~\cite{raNonGaussianQuantumStates2020a,walschaersNonGaussianQuantumStates2021,zhangMaximalEntanglementIncrease2022} of light by utilizing such a PDC-based nonlinear quantum interference. Two-photon interference serves as the basic element for large scale photonic quantum systems. Thus, by generalizing to multiple photons and multiple modes~\cite{jabbourMultiparticleQuantumInterference2021,salazarLinearnonlinearDualityCircuit2023,bezerraFamiliesBosonicSuppression2023a}, we anticipate that two-photon nonlinear quantum interference may have enormous applications in optical quantum computation and quantum information processing.




\vspace{-6px}

\bibliography{ref}

\section*{Methods}
\noindent \textit{Determination of $P_1$ from multi-channel coincidence measurement.} For an arbitrary state with photon number distribution $P_n$, the probability that $m$ detectors respond simultaneously is 
\begin{equation}
	\begin{aligned}
		C_m&=\sum_{n=m}^\infty P_n*P^{(n)}_m \\
		&=\sum_{n=m}^\infty P_n\sum_{r=0}^m(-1)^r\frac{m!}{r!(m-r)!}(1-r\eta)^n \\
	\end{aligned}
    \label{eq:C_m}
\end{equation}
Since the detection efficiency $\eta$ can be measured experimentally, there is a deterministic relationship between $C_m$ and $P_n$. Therefore, $P_1$ also can be calculated by measured $C_m$ (see Supplementary Information).
In realistic experiment, the number of detectors is limited. For example, $m$ detectors enable the measurement of $C_1,C_2,\cdots,C_m$. We select $C_1,C_2,\cdots,C_5$ to approximate $P_1$, which is the upper bound of real $P_1$. This approximation is valid when  $g\lesssim 1.2$. In this case, the response probability of $m$ detectors, $C_m$, is negligible small for $m\geq 6$, i.e., the $n~(n\geq 6)$-photon component is insignificant. 

To resolve $P_1$ when $g> 1.2$, we create a model characterized by parameters -- gain $g$, mode match and detection efficiency to describe two-photon CJ nonlinear interference experiments (see Supplementary Information). All the parameters are fitted in auxiliary experiments, where the input states for PPKTP3 are changed. Then we substitute the fitted parameters into the model to calculate the $P_1$ of the output state for two-photon nonlinear interference experiment when $g\approx 2$. The deduced $P_1$ from the fitted model agrees well with the approximated $P_1$ from multi-channel coincidence measurement for experiments of $g\approx 1.2$, confirming the validity of our fitted model.

\bigskip
\bigskip
\noindent\textbf{Data availability}\\ The data supporting the study and figures are available upon request, which should be addressed to Chao Chen.

\medskip
\noindent\textbf{Acknowledgements}\\ We thank the team of \textit{Jiuzhang} quantum computer for technical support of vital importance. This work was supported financially by the National Natural Science Foundation of China (Grants No. 12234009, No. 12274215 and No. 12304398); the National Key R\&D Program of China (Grants No. 2019YFA0308700 and No. 2020YFA0309500); the Innovation Program for Quantum Science and Technology (Grant No. 2021ZD0301400); the Program for Innovative Talents and Entrepreneurs in Jiangsu; Key R\&D Program of Jiangsu Province (Grant No. BE2023002); the Key R\&D Program of Guangdong Province (Grant No. 2020B0303010001); the Natural Science Foundation of Jiangsu Province (Grant No. BK20220759); the China Postdoctoral Science Foundation (2023T160297). MGJ acknowledges support by the Fonds de la Recherche Scientifique – FNRS, as well as by the Carlsberg Foundation. NJC acknowledges support from the European Union under project
ShoQC within the ERA-NET Cofund in Quantum Technologies (QuantERA) program, as well as from the Fonds de la Recherche Scientifique – FNRS under project CHEQS within the Excellence of
Science (EOS) program.

\medskip
\noindent\textbf{Author contributions}\\ Xi-Lin Wang and Hui-Tian Wang designed the research and supervised the project; Chao Chen, Shu-Tian Xue, Yu-Peng Shi, Xi-Lin Wang and Hui-Tian Wang performed the experiment and implemented the numerics; Nicolas J. Cerf, Chao Chen, Shu-Tian Xue, Xi-Lin Wang and Hui-Tian Wang analyzed the results and wrote the manuscript; all authors discussed the results and reviewed the manuscript.


\medskip
\noindent\textbf{Competing interests}\\ The authors declare no competing interests.

\medskip
\noindent\textbf{Additional information}\\ Further material and details on the calculations are provided in Supplementary Information.

\end{document}


\title{Supplementary Information for:\vspace{.3cm}\\~``Two-particle quantum interference in a nonlinear optical medium:\\ a witness of timelike indistinguishability"}
\author{Chao Chen, Shu-Tian Xue, Yu-Peng Shi, Jing Wang, Zi-Mo Cheng, Pei Wan, Zhi-Cheng Ren, Michael G. Jabbour, Nicolas J. Cerf, Xi-Lin Wang, Hui-Tian Wang}

\maketitle

\tableofcontents

\newpage

\section{Two-photon Linear and Nonlinear Interference}
\subsection{Linear Hong-Ou-Mandel interference}
Hong-Ou-Mandel (HOM) interference describes two indistinguishable single photons interfering at a beam splitter (BS). The interference gives rise to the two photons bunching at either of the two output modes of the beam splitter, and the probability of finding one single photon in each of the two output modes of the beam splitter vanishes.  The output state of HOM interference can be written as
\begin{equation}
\ket{\psi}^\text{HOM}=U_{T}^{\text{BS}}\ket{1,1}=U_{T}^{\text{BS}}\hat{a}^\dagger \hat{b}^\dagger \ket{0,0},
\end{equation}
with $T$ the transmittance of the BS, $\hat{a}$ and $\hat{b}$ the two modes of the BS. $U_{T}^{\text{BS}}$ is the unitary operation of a BS, namely
\begin{equation}
U_{T}^{\text{BS}}=\text{exp}[\theta (\hat{a}^\dagger b-\hat{b}^\dagger a)],T=\cos^2{\theta}.
\end{equation}
In the Heisenberg picture, the mode operators $\hat{a}$ and $\hat{b}$ act as
\begin{align}
U_{T}^{\text{BS}}\hat{a}^\dagger {U_{T}^{\text{BS}}}^\dagger&=\hat{a}^\dagger\cos{\theta}-\hat{b}^\dagger\sin{\theta},
\label{eq:heiserbergbs1}\\
U_{T}^{\text{BS}}\hat{b}^\dagger {U_{T}^{\text{BS}}}^\dagger&=\hat{a}^\dagger\sin{\theta}+\hat{b}^\dagger\cos{\theta}. 
\label{eq:heiserbergbs2}
\end{align}
With Eq.~\ref{eq:heiserbergbs1} and \ref{eq:heiserbergbs2}, we rewrite the output state,
\begin{align}
\ket{\psi}^\text{HOM}&=U_{T}^{\text{BS}}\hat{a}^\dagger {U_{T}^{\text{BS}}}^\dagger U_{T}^{\text{BS}}\hat{b}^\dagger {U_{T}^{\text{BS}}}^\dagger U_{T}^{\text{BS}}\ket{0,0}\\
&=(\hat{a}^\dagger\cos{\theta}-\hat{b}^\dagger\sin{\theta})(\hat{a}^\dagger\sin{\theta}+\hat{b}^\dagger\cos{\theta})\ket{0,0}\\
&=(\cos{\theta}\sin{\theta}\hat{a}^{\dagger^2}+\cos^2{\theta}\hat{a}^\dagger\hat{b}^\dagger-\sin^2{\theta}\hat{b}^\dagger\hat{a}^\dagger-\sin{\theta}\cos{\theta}\hat{b}^{\dagger^2})\ket{0,0}.
\end{align}
To understand HOM interference, we calculate the probability of outputting one photon in each mode,
\begin{align}
    P_{1,1}^\text{HOM}&=\left|\bra{1,1}\cos^2{\theta}\hat{a}^\dagger\hat{b}^\dagger-\sin^2{\theta}\hat{b}^\dagger\hat{a}^\dagger\ket{0,0}\right|^2
     \label{eq:p11HOM1}\\
    &=(\cos^2{\theta}-\sin^2{\theta})^2
    \label{eq:p11HOM2} \\
    &=(2\cos^2{\theta}-1)^2=(2T-1)^2
\end{align}
From Eq.~\ref{eq:p11HOM1} to Eq.~\ref{eq:p11HOM2}, the commutation relation for bosons is used, namely,
\begin{equation}
    \left[\hat{a}^\dagger,\hat{b}^\dagger\right]=0.
\end{equation}
For a beam splitter with transmittance $T=1/2$, the probability of detecting coincidence at the output of the beam splitter drops to zero. This gives rise to the well-known HOM dip phenomenon when tuning the time delay of one input photon with respect to another single photon. In HOM interference, the beam splitter couples the $\hat{a}$ and $\hat{b}$ modes linearly. After the transformation of the beam splitter, the photon number is conserved.

\subsection{Two-photon Cerf-Jabbour quantum interference in nonlinear optical medium}
If we replace the beam splitter by a nonlinear medium where parametric down conversion (PDC) is allowed, the HOM interference can be extended to nonlinear regime as a consequence of the duality between the BS and PDC under partial time reversal~\cite{PTR}. In the main text, we name this two-photon nonlinear quantum interference after N. J. Cerf and M. G. Jabbour as Cerf-Jabbour (CJ) interference~\cite{cerfTwobosonQuantumInterference2020}. Likewise, firstly we give the unitary of parametric down conversion process,
\begin{equation}
    U_g^{\text{PDC}}=\text{exp}[r(\hat{a}_\text{H}^\dagger\hat{a}_\text{V}^\dagger-\hat{a}_\text{H}\hat{a}_\text{V})], g=\cosh^2{r},
\end{equation}
where $g$ is the parametric gain and $r$ is the squeezing parameter. The output state of CJ nonlinear interference can be written as
\begin{equation}
    \ket{\psi}^\text{CJ}=U_g^{\text{PDC}}\ket{1,1}=U_g^{\text{PDC}}\hat{a}_\text{H}^\dagger {U_g^\text{PDC}}^\dagger U_g^{\text{PDC}}\hat{a}_\text{V}^\dagger {U_g^\text{PDC}}^\dagger U_g^{\text{PDC}}\ket{0,0}.
    \label{eq:psiNHOM}
\end{equation}
In the Heisenberg picture, the modes operators in Eq.~\ref{eq:psiNHOM} transform as,
\begin{align}
   U_g^{\text{PDC}}\hat{a}_\text{H}^\dagger {U_g^{\text{PDC}}}^\dagger&=\hat{a}_\text{H}^\dagger\cosh{r}-\hat{a}_\text{V}\sinh{r},
   \label{eq:heiserbergpdc1}\\
U_g^{\text{PDC}}\hat{a}_\text{V}^\dagger{U_g^{\text{PDC}}}^\dagger&=-\hat{a}_\text{H}\sinh{r}+\hat{a}_\text{V}^\dagger\cosh{r}. 
\label{eq:heiserbergpdc2} 
\end{align}
If we act $U_g^\text{PDC}$ on a vacuum state, a two-mode squeezed vacuum state is obtained, namely,
\begin{equation}
    U_g^\text{PDC}\ket{0,0}=\frac{1}{\cosh{r}}\sum_{n=0}^\infty \tanh^n{r} \ket{n,n}.
    \label{eq:two-mode-sq}
\end{equation}
By making the substitutions shown in Eqs.~\ref{eq:heiserbergpdc1}, \ref{eq:heiserbergpdc2} and \ref{eq:two-mode-sq}, we get the probability of outputting one pair photon after CJ interference as,
\begin{align}
    P_{1,1}^\text{CJ}&=\left|\bra{1,1}(\cosh^2{r} \frac{1}{\cosh{r}} \hat{a}_\text{H}^\dagger\hat{a}_\text{V}^\dagger\ket{0,0}-\cosh{r}\sinh{r}\frac{\tanh{r}}{\cosh{r}}(\hat{a}_\text{H}^\dagger\hat{a}_\text{H}+\hat{a}_\text{V}\hat{a}_\text{V}^\dagger)\ket{1,1} \right. \\
    & \left. \quad \quad +\sinh^2{r} \frac{\tanh^2{r}}{\cosh{r}}\hat{a}_\text{H}\hat{a}_\text{V}\ket{2,2})\right|^2\\
    &=\left(\cosh{r}-\frac{3\sinh^2{r}}{\cosh{r}}+\frac{2\sinh^4{r}}{\cosh^3{r}}\right )^2\\
    &=\frac{(2-\cosh^2{r})^2}{\cosh^6{r}}=\frac{(2-g)^2}{g^3}.
    \label{eq:p11nhom}
\end{align}
Therefore, tuning the parametric gain $g$ changes the depth of the interference. When $g=2$, $P_{1,1}^\text{CJ}$ vanishes because of the completely destructive interference of the one-photon-pair state~\cite{cerfTwobosonQuantumInterference2020}.

\begin{figure}
\centering
\includegraphics[width=\textwidth]{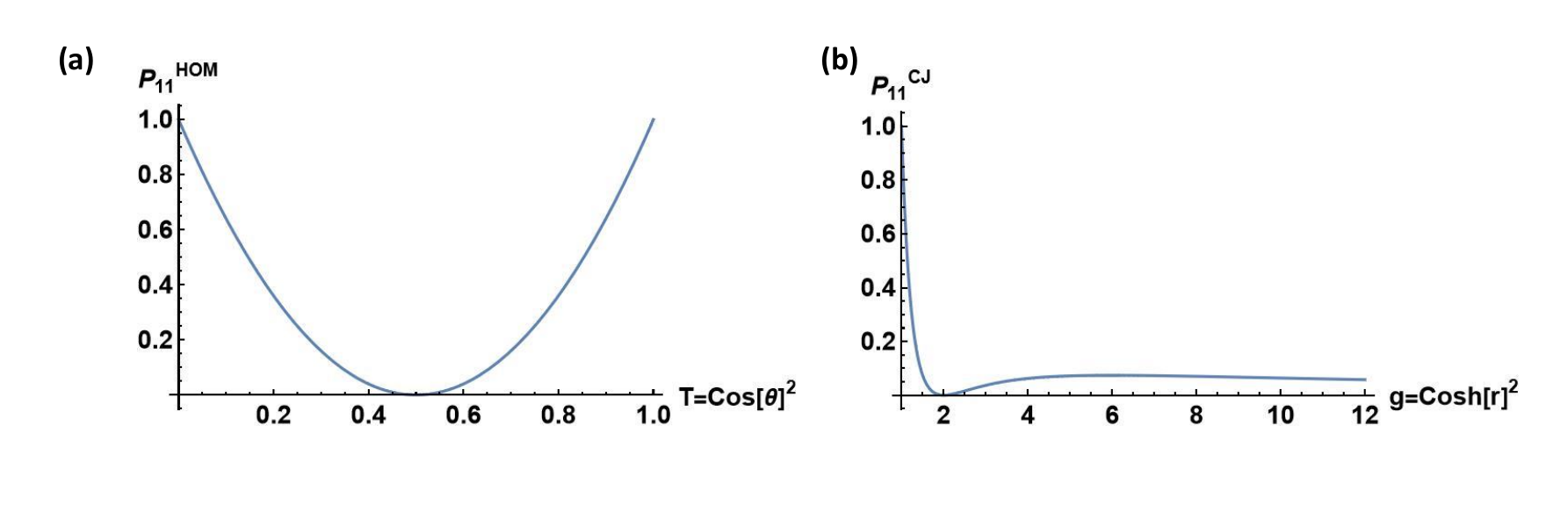}
\caption{ (a)~Probability of outputting one photon in each mode  versus the transmittance $T$ of the beam splitter in HOM interference.
(b)~Probability of outputting one pair of photons versus the parametric gain $g$ of PDC in two-photon CJ nonlinear interference.}
\label{fig:HOM depth}
\end{figure}

\begin{figure}
\centering
\includegraphics[width=0.6\textwidth]{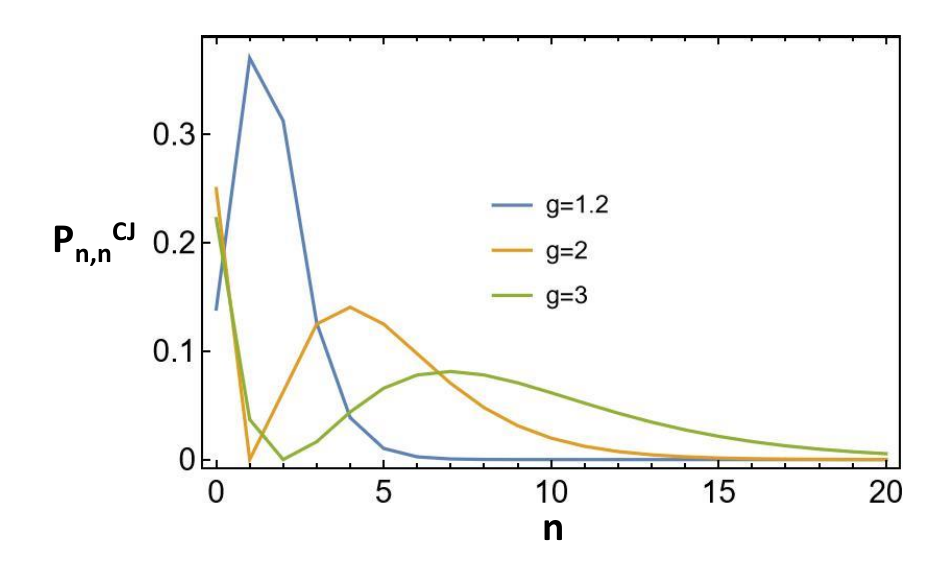}
\caption{Probability of outputting $n$ photon pairs after CJ nonlinear interference of different $g$.}
\label{fig:pn_distribution}
\end{figure}

As shown in Fig.~\ref{fig:HOM depth}, $P_{1,1}^\text{HOM}$ is symmetric with respect to $T=0.5$. However, $P_{1,1}^\text{CJ}$ behaves dramatically different, reflecting the nonlinear nature of two-photon interference in PDC. Notably, photon number is not conserved in nonlinear quantum interference. Besides outputting one photon pair, there are probabilities of outputting two photon pairs, three photon pairs and so on. Generally, the whole output state reads,
\begin{equation}
   U_g^\text{PDC}\ket{1,1}=\sum_{n=0}^\infty\frac{(\sinh{r})^{n-1}}{(\cosh{r})^{n+2}}(n-\sinh^2{r})\ket{n,n}.
\end{equation}
In Fig.~\ref{fig:pn_distribution}, we show the output photon number distributions of CJ interference of difference parametric gain. When $g=3$, the probability of outputting two photon pairs falls down to zero, corresponding to four-photon destructive interference in the PDC. More generally, when gain $g$ is an integer larger than 2,  the component of outputting $(g-1)$ photon pairs vanishes~\cite{cerfTwobosonQuantumInterference2020}. 

\section{Parametric Down Conversion (PDC) Source}
\label{sec:2}
In this section, we introduce the properties of our PDC source and the methods to reach uncorrelated joint spectrum. The PDC process happens in a type-II periodically poled potassium titanyl phosphate (PPKTP) crystal, where pairs of photons of 1558 nm are generated collinearly when the PPKTP crystal is pumped by a horizontally polarized laser light of 779 nm. There are three PPKTP crystals in our experiment as shown in Fig.~2 in the main text. The PPKTP1 and PPKTP2 are used to generate heralded single photons of horizontal and vertical polarization, respectively. The heralded single photons should be matched with the PDC modes of PPKTP3 in all degrees of freedom, such as spectrum, path and time. 

\begin{figure}
\centering
\includegraphics[width=0.6\textwidth]{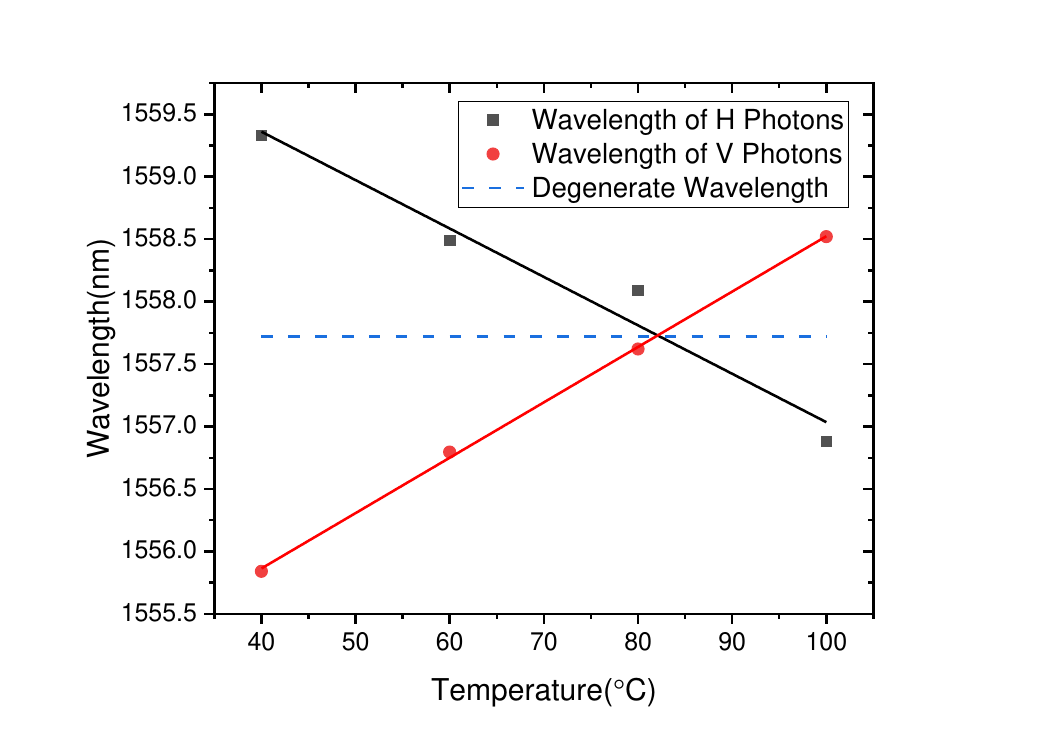}
\caption{Central wavelength of the photon pairs dependent on the temperature of the PPKTP crystal. The temperature is controlled to keep the wavelengths of photon pairs degenerate.}
\label{fig:tem_control}
\end{figure}

To ensure the spectral indistinguishability of single photons, PPKTP1, PPKTP2 and PPKTP3 are all temperature controlled. As shown in Fig.~\ref{fig:tem_control}, we tune the temperature so that the horizontally polarized and vertically polarized photons are both at degenerate central wavelength, which is double the wavelength of the pump laser. Another important thing is to make the spectrum of the heralded single photons uncorrelated with that of the photons detected for trigger. The original joint spectrum $S(w_i,w_s)$ has side lobes due to the sinc function for phase matching spectrum. 

\begin{figure}
\centering
\includegraphics[width=0.6\textwidth]{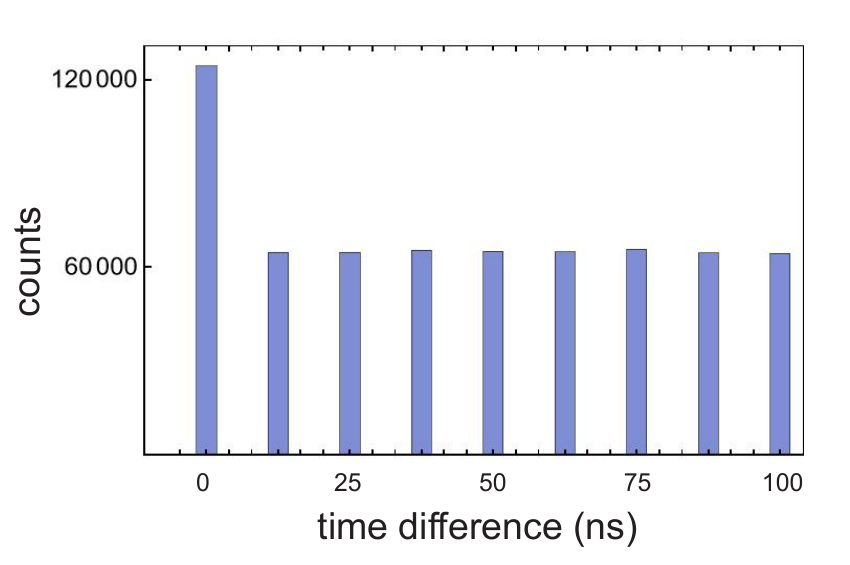}
\caption{Experimentally measured unheralded second-order correlation histogram of H-polarized photons after 15 nm filters. By compared the first bar of zero-time-delay coincidence counts with pulse-period-delay coincidence counts, we have $g^2(0)=1.911$. The average photon number for the $g^2(0)$ measurement is 0.021. The corresponding spectral purity of the filtered photons is about 0.93.}
\label{fig:purity}
\end{figure}

To quantify the frequency correlation, we use purity, which can be calculated by doing Schmidt decomposition for $S(w_i,w_s)$. The Schmidt decomposition of $S(w_i,w_s)$ is expressed as,
\begin{equation}
    S(w_i,w_s)=\sum_j c_j \psi_j (w_i) \phi_j(w_s)
    \label{eq:schimdt}
\end{equation}
where ${\psi_j(w)}$ and ${\phi_j(w)}$ is a biorthogonal system and $\sum_j c_j^2=1$. When tracing out one of the two down-conversion modes, the spectral purity of the other mode is 
\begin{equation}
\text{tr}_i(\rho^2_{s,i})=\text{tr}_s(\rho^2_{s,i})=\sum_j c_j^4.
\label{eq:purity}
\end{equation}
When the purity is 1, the joint spectrum is uncorrelated.

In the experiment, we filter out the side lobes by using 15 nm filters to promote the spectral purity. The joint spectrum after introducing a 15 nm filter for the horizontally polarized photons is shown in Fig.~2(b) in the main text. We perform unheralded second-order correlation measurement to obtain $g^{(2)}(0)$ in a Hanbury Brown and Twiss (HBT) type setup. If $g^{(2)}(0)$ is measured by ideal detectors, the purity of the detected photons is $\mathbb{P}=g^{(2)}(0)-1$. However, threshold single-photon detectors are employed instead of ideal detectors. Following Ref. \cite{zhongQuantumComputationalAdvantage2020}, we measure $g^{(2)}(0)$ and average photon number to get corrected purity $\mathbb{P}$. After the side lobes are filtered out, the typical measured $g^{(2)}(0)$ is shown in Fig.~\ref{fig:purity}. The average purity of the three PPKTP sources is about 0.92.

\section{Detection Efficiency and Count Rate Correction}
\label{sec:3}
Detection efficiency $\eta$ can be calculated from the coincidence measurement of H-polarized and V-polarized mode. We use $N_H(V)$ to represent the count rate of the H (V)-polarized mode, and $N_{HV}$ to represent the coincidence count rate of H- and V-polarized modes. Then the overall detection efficiency of V (H)-polarized photons is,
$\eta_{V(H)}=\nicefrac{N_{H(V)}}{N_{HV}}$.

Due to the existence of dead time of the detector and the fact that the second photon may arrive in dead time of the first photon, the actual count rate obtained from the measurement is reduced.
Especially when the brightness of  photons at 1558 nm is high in this experiment, a significant reduction of count rate occurs.

Superconducting nanowire single photon detectors have an average dead time $t_d\simeq 18$ ns, and the pump laser has pulse interval $t_0=12.5$ ns. Deadtime covers $n_d=[t_d/t_0]=1$ pulse.
Assuming the probability of the presence of photons in a single pulse is $p$, the expectation of the time interval between two responses of a detector is
\begin{equation}
	t_E = n_d t_0+\sum_{n=1}^\infty n t_0\times p\Big(1-p\Big)^{n-1}
	= t_0\Big(n_d+\frac{1}{p}\Big).
\end{equation}

The measured response probability of each single detector is
\begin{equation}
	p^*=\frac{t_0}{t_E}=\frac{p}{1+n_dp}.
\end{equation}
Accordingly, the genuine probability $p$ of having photons in a single pulse can be calculated
\begin{equation}
	p=\frac{p^*}{1-n_d p^*}.
\end{equation}
Then, the consequent attenuation of the count rate due to the dead time is
\begin{equation}
	\lambda(p)=\frac{p^*}{p}=\frac{1}{1+n_dp}=1-n_d p^*.
\end{equation}
We divide the measured count rate by the coefficient $\lambda(p)$ to get the corrected count rate.

\section{Analysis of Photon Number Statistics}
\label{sec:4}
Different states lead to different responses of the detector array.
Detectors' coincidence counts are related to photon number distributions, and vice versa.
Here, we introduce how to resolve the photon number distribution by measuring the coincidence counts of multiple detectors.

\subsection{Response probability of a Fock state $\ket{n}$}
Consider a $n$-photon state $\ket{n}$ is injected into an array of detectors. The detection efficiency of the $k$-th detector is represented by $\eta_{k}$. Regardless of whether the other detectors respond or not, the response probability of the $k$-th detector $P^{(n)}_k$ to the state $\ket{n}$ can be expressed as
\begin{equation}
	P^{(n)}_k = 1-(1-\eta_k)^n.
\end{equation}
Similarly, the probability of the $k_1$-th and $k_2$-th detectors response simultaneously is
\begin{equation}
	P^{(n)}_{k_1,k_2} = 1-(1-\eta_{k_1})^n-(1-\eta_{k_2})^n+(1-\eta_{k_1}-\eta_{k_2})^n.
\end{equation}

According to the principle of inclusion and exclusion, it can be derived that the probability of $m$ ($m\leq n$) detectors responding simultaneously is
\begin{equation}
	P^{(n)}_{k_1,k_2,\cdots,k_m}=\sum_{r=0}^m(-1)^r\sum_{\{l_1,l_2,\cdots,l_r\}\in\{k_1,k_2,\cdots,k_m\}}(1-\sum_{l_1,l_2,\cdots,l_r}\eta_{l})^n.
\end{equation}
$\{l_1,l_2,\cdots,l_r\}$ is chosen from $\{k_1,k_2,\cdots,k_m\}$.

In our experiments, we tune the overall detection efficiency of each detector to be equal by rotating the half wave plates before the polarization beam splitters in the photon number analysis module. Thus, it's a practical approximation that $\eta_1\simeq\eta_2\cdots\simeq\eta_k\cdots=\eta$. The response probability of arbitrary $m$ detectors is
\begin{equation}
	P^{(n)}_m=\sum_{r=0}^m(-1)^r\frac{m!}{r!(m-r)!}(1-r\eta)^n.
\end{equation}
Notably, $P^{(n)}_m$ corresponds to the $m$-fold coincidence counts in the experiment. It is different from the case that only $m$ detectors respond.

\subsection{Response probability of a state with photon number distribution $P_n$}
For an arbitrary state with the probability of having $n$ photons represented by $P_n$, the response probability of $m$ detectors 
\begin{equation}
	\begin{aligned}
		C_m&=\sum_{n=m}^\infty P_n*P^{(n)}_m \\
		&=\sum_{n=m}^\infty P_n\sum_{r=0}^m(-1)^r\frac{m!}{r!(m-r)!}(1-r\eta)^n. \\
	\end{aligned}
    \label{eq:C_m}
\end{equation}
This equation expands in the following form:
\begin{displaymath}
	\begin{array}{rcrrrrrrrrr}
	C_1&=&\eta^1[&P_1+&(2-\eta)P_2+&(3-3\eta+\eta^2)P_3+&(4-6\eta+4\eta^2-\eta^3)P_4+&\cdots] \\
	C_2&=&\eta^2[&&2P_2+&(6-6\eta)P_3+&(12-24\eta+14\eta^2)P_4+&\cdots] \\
	C_3&=&\eta^3[&&&6P_3+&(24-36\eta)P_4+&\cdots] \\
	C_4&=&\eta^4[&&&&24P_4+&\cdots] \\
	\cdots \\
	\end{array}
\end{displaymath}

Since the detection efficiency $\eta$ can be measured experimentally, there is a deterministic relationship between $C_m$ and $P_n$.
However, infinite number of detectors are required to calculate the complete photon number distribution $P_n$ via measured $m$-fold coincidence probability mathematically, because $n$ is infinite in principle.

\subsection{Approximate calculation of $P_1$}
\label{sec:p1}
For $m$-fold coincidence probability, arbitrary $n$-photons term which satisfies $n \geqslant m$ contributes.
Although the response probability of $m$-detector coincidence for $n$-photons term (denoted as $P^{(n)}_m$ above) increases with $n$, the $n$-photons probability $P_n$ decreases rapidly for an arbitrary state with finite average photon number as long as $n$ is sufficiently large. Thus, the overall contribution of $n$-photons component decreases gradually after a certain value of $n$.
We may find a suitable position for truncation, so that the contribution of neglected terms with more photons has little effect in the calculation of $P_1$.

According to Eq.~\ref{eq:C_m}, the probability of only having one photon can be written strictly as
\begin{equation}
	\begin{aligned}
		P_1=&~\frac{1}{\eta^1}C_1
		-\frac{2-\eta}{2!\eta^2}C_2
		+\frac{3-6\eta+2\eta^2}{3!\eta^3}C_3
		-\frac{4-18\eta+22\eta^2-6\eta^3}{4!\eta^4}C_4 \\
		&~+\frac{5-40\eta+105\eta^2-100\eta^3+24\eta^4}{5!\eta^5}C_5
		-\frac{6-75\eta+340\eta^2-675\eta^3+548\eta^4-120\eta^5}{6!\eta^6}C_6 \\
		&~+\sum_{m=7}^\infty \alpha_m(\eta) ~C_m. \\
	\end{aligned}
    \label{eq:P1}
\end{equation}
We rewrite the last term in Eq.~\ref{eq:P1} by photon number distribution $P_n$ as
\begin{equation}
	\sum_{m=7}^\infty \alpha_m(\eta) ~C_m
	=\sum_{n=7}^\infty \beta_n(\eta) ~P_n.
\end{equation}
$\alpha_m(\eta)$ and $\beta_n(\eta)$ are parameters that can be derived from  Eq.~\ref{eq:C_m}.

For experiments with the power of pump laser less than 700 mW (corresponding $g\sim1.2$ ), the multi-photon component $P_n$ of the output state is negligible when $n\geqslant7$. Hence the $C_m (m\geqslant7)$ originates from the contribution of $P_n$ with $n\geqslant m$ is also insignificant. In this case, the last term in Eq.~\ref{eq:P1} can be neglected, and we use $C_{1-6}$ to approximately calculate $P_1$.

Using $C_1,C_2,\cdots,C_6$ measured from 6 detectors, we can calculate an approximation of the probability $P_1$,
\begin{equation}
	\begin{aligned}
		P_1\simeq&~P_1^\mathrm{6-detectors} \\
		=&~\frac{1}{\eta^1}C_1
		-\frac{2-\eta}{2!\eta^2}C_2
		+\frac{3-6\eta+2\eta^2}{3!\eta^3}C_3
		-\frac{4-18\eta+22\eta^2-6\eta^3}{4!\eta^4}C_4 \\
		&~+\frac{5-40\eta+105\eta^2-100\eta^3+24\eta^4}{5!\eta^5}C_5
		-\frac{6-75\eta+340\eta^2-675\eta^3+548\eta^4-120\eta^5}{6!\eta^6}C_6. \\
	\end{aligned}
\end{equation}

The approximate $P_1$ using 6 detectors satisfies $P_1^\mathrm{6-detectors} < P_1$. $C_6$ has low count rates, and therefore has larger statistic errors.
In order to characterize $P_1$ with smaller measurement uncertainties than $P_1^\mathrm{6-detectors}$, in the main text we select $C_1,C_2,\cdots,C_5$ to calculate $P_1$ by the following approximation,
\begin{equation}
	\begin{aligned}
		P_1\simeq&~P_1^\mathrm{5-detect} \\
		=&~\frac{1}{\eta^1}C_1
		-\frac{2-\eta}{2!\eta^2}C_2
		+\frac{3-6\eta+2\eta^2}{3!\eta^3}C_3
		-\frac{4-18\eta+22\eta^2-6\eta^3}{4!\eta^4}C_4 \\
		&~+\frac{5-40\eta+105\eta^2-100\eta^3+24\eta^4}{5!\eta^5}C_5.\\
	\end{aligned}
\end{equation}
 $P_1^\mathrm{5-detectors}$ is the upper bound of the real $P_1$, i.e., $P_1^\mathrm{5-detectors}> P_1$. 
\begin{figure}
\centering
\includegraphics[width=1\textwidth]{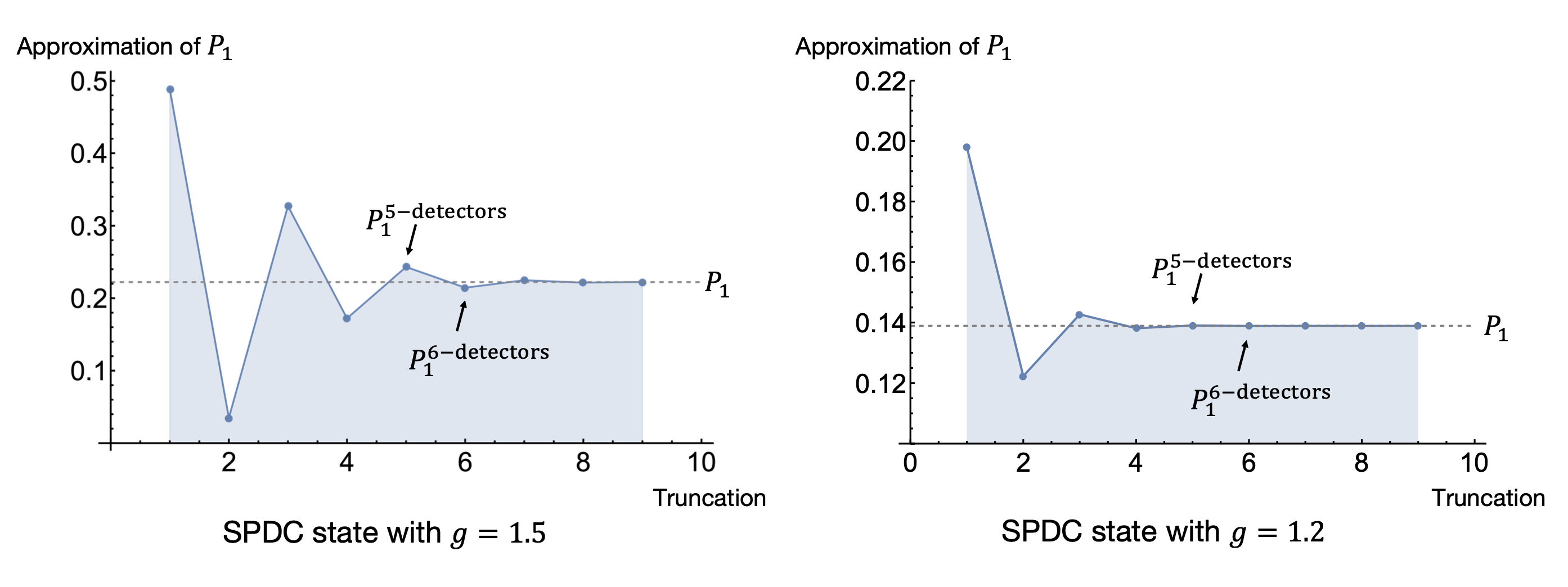}
\caption{Calculated $P_1$ with truncation at various $m$ for the two-mode squeezed vacuum state obtained by acting $U_{g}^{\text{PDC}}$ of $g=1.5$ and $g=1.2$ on a vacuum state. The dashed line represents the exact value of $P_1$.}
\label{fig:P_1_truncation}
\end{figure}

\begin{figure}
\centering
\includegraphics[width=0.6\textwidth]{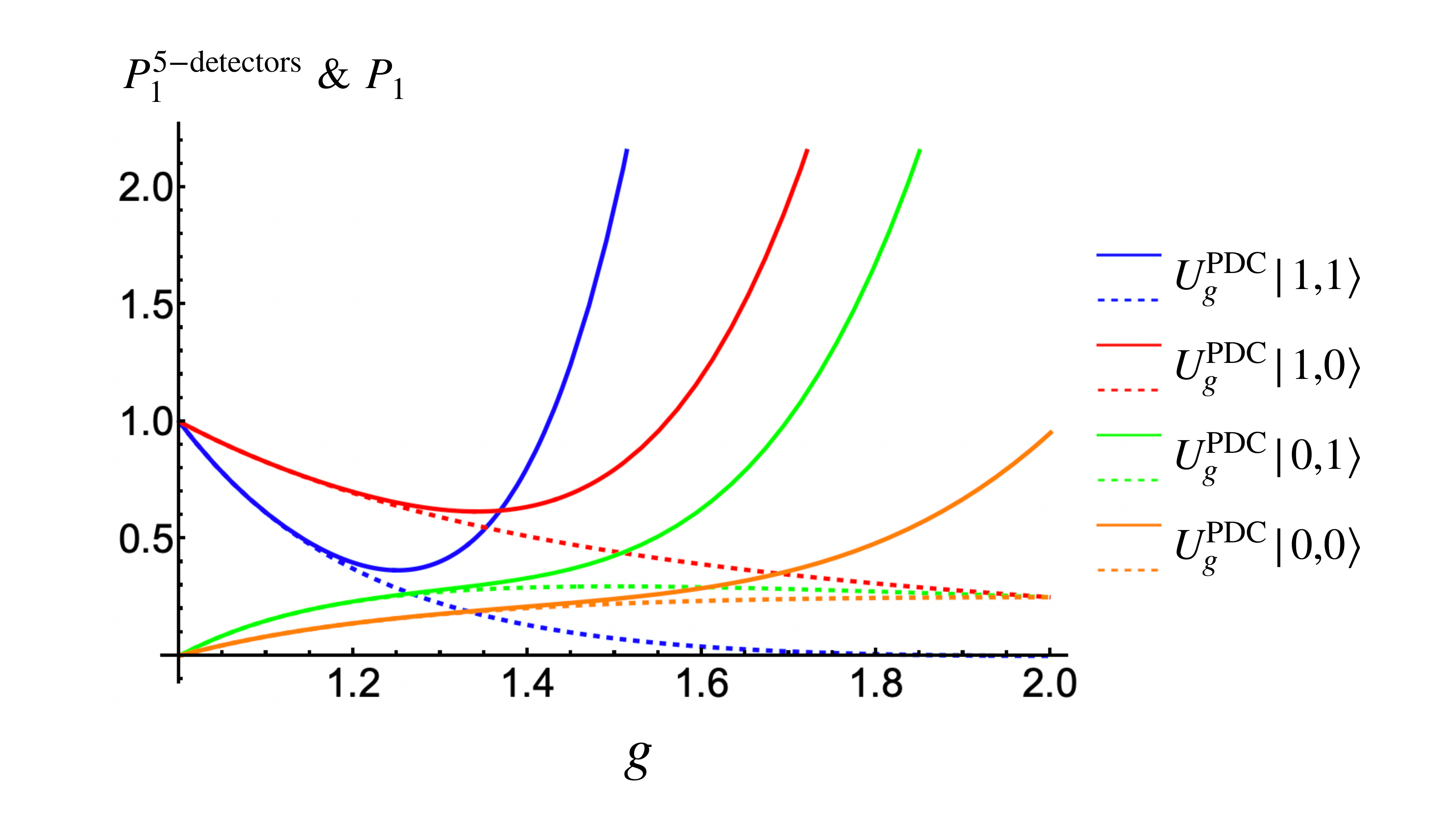}
\caption{Comparison of calculated $P_1^\mathrm{5-detectors}$ and actual $P_1$ versus the gain $g$. Solid lines represent $P_1^\mathrm{5-detectors}$ and dashed lines represent the real $P_1$. The different colors correspond to the output states when the PDC unitary operation is applied on different input states.}
\label{fig:valid}
\end{figure}

In Fig.~\ref{fig:P_1_truncation}, we show the calculated $P_1$ with truncated $C_m$ when the parametric gain $g=1.5$ and $g=1.2$. $P_1^\mathrm{5-detectors}$ fits well with the theoretical $P_1$ in the condition of $g=1.2$. In this case, $C_m$, the response probability of $m$ detectors, is negligible small for $m\geq 6$, i.e., the $n~(n\geq 6)$-photon component is insignificant. Fig.~\ref{fig:valid}, we further show the deviation of $P_1^\mathrm{5-detectors}$ from the real $P_1$ against the parametric gain $g$ of the PDC for various input states. Our simulation proves the feasibility of approximating $P_1$ from the coincidence measurement of a few numbers of detectors when $g$ is small. However, when $g>1.2$, $P_1^\mathrm{5-detectors}$ starts to deviate from the real $P_1$. That's why we restrict the $g\approx 1.2$ in the experiments to get the results shown in Fig.~3 of the main text.

\section{Deducing $P_1$ from A Fitted Experimental Model}
\label{sec:deducingP1}
In section \ref{sec:p1}, we introduce how to get approximate $P_1$ directly from measurable multi-channel coincidence when the parametric gain $g$ is small. Here, we will give a model functioned by a few parameters to reliably describe the two-photon nonlinear quantum interference process in PDC. First, all the parameters are fitted from the measured results of auxiliary experiments, where the input state for PPKTP3 are changed. Then we substitute the fitted parameters into the model of two-photon CJ interference experiment to derive the photon number distribution of the output state, thus avoiding the direct measurement of the photon number distribution $P_n$. 
We verify this model of two-photon CJ interference experiment by comparing the derived $P_1$ with the measured $P_1^\mathrm{5-detectors}$ of experiment at small $g$. In Fig.~4(a) of the main text, we show that the measured $P_1^\mathrm{5-detectors}$ agree well with the deduced $P_1$ calculated from our fitted experimental model for $g=1.21$. Therefore, this verified method can be applied to cases of $g>1.2$ where the approximation of $P_1$ from directly measured $m$-fold coincidence is invalid.

\subsection{Measurement of parametric gain of PDC}
\label{sec:g}
When the heralded single photons are blocked, spontaneous parametric down conversion (SPDC) happens in PPKTP3. The output state of PPKTP3 is a two-mode squeezed vacuum state described by parametric gain $g$
\begin{equation}
	\ket{\psi}^\mathrm{SPDC}=U^\mathrm{PDC}_g\ket{0,0}=\frac{1}{\sqrt{g}}\sum_{n=0}^\infty\Big(\frac{g-1}{g}\Big)^{n/2}\ket{n,n}.
\end{equation}
The corresponding photon number distribution is
\begin{equation}
	P_n^\text{SPDC}=\frac{(g-1)^n}{g^{n+1}}.
\end{equation}
Substituting $P_n^\mathrm{SPDC}$ into Eq.~\ref{eq:C_m}, we have $C_m$ to be a function of $g$ and $\eta$. Since $\eta$ is known (see \ref{sec:3}), $C_m$ is only dependent on $g$.

\begin{figure}
\centering
\includegraphics[width=0.6\textwidth]{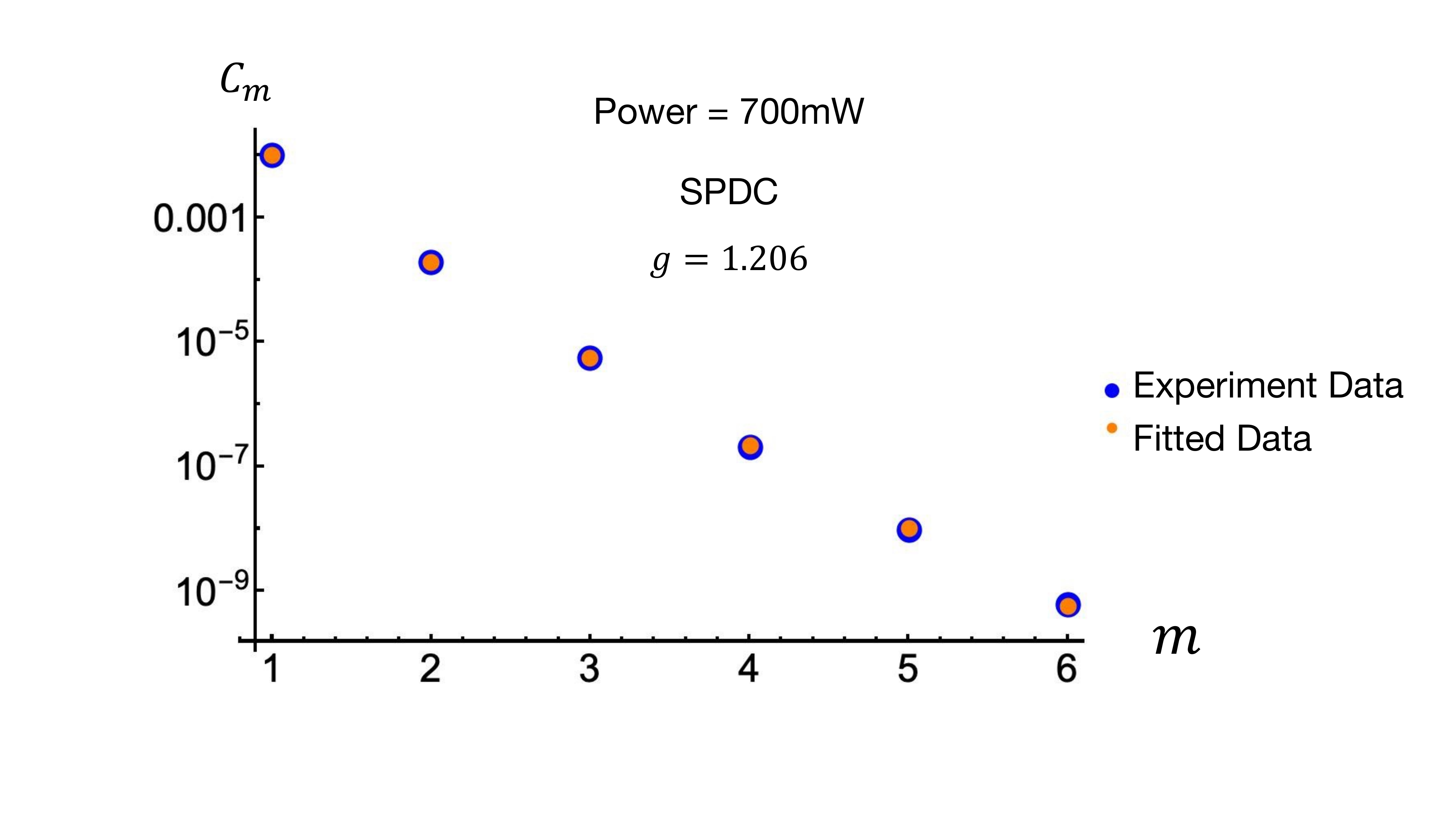}
\caption{Measured $m$-fold coincidence $C_m$ of the H-polarized output mode and the theoretically calculated $C_m$ parameterized by the fitted $g$ for a two-mode squeezing vacuum state.}
\label{fig:fit_00}
\end{figure}

We fit the measured $m$-fold coincidence probability $C_1,C_2,\cdots,C_6$ with the ones calculated by \ref{eq:C_m} to get the parametric gain $g$.
In Fig.~\ref{fig:fit_00}, we show that the measured $C_m$ agrees 
well with the theoretically calculated $C_m$ parameterized by the fitted gain $g$. 

\subsection{Measuring the overlap of the heralded single photons and the modes of PPKTP3}
\label{sec:mathcal{O}}
We have discussed how to get the fitted gain $g$ by coincidence measurement of the output state of SPDC in last section. Here we consider the case where one single photon is initially injected, which can be written as  $|1,0\rangle$ or $|0,1\rangle$.
After implement the $U^{\text{PDC}}$ on the initial state, the output photon-number distribution of the H-polarized mode can be written as,
\begin{equation}
	\begin{aligned}
		P_n^{\ket{1,0}}&=\frac{(g-1)^{n-1}}{g^{n+1}}n \\
		P_n^{\ket{0,1}}&=\frac{(g-1)^n}{g^{n+2}}(n+1) \\
	\end{aligned}
\end{equation}

However, such a perfect initial state is almost impossible to realize in experiment. It is difficult to perfectly overlap the mode of the incident photons with the modes of the PDC process in all degrees of freedom, such as time, space, and spectrum. The overlap of the incident H (V)-polarized photon with the H (V)-polarized PDC mode is denoted by parameter $\mathcal{O}_1$ ($\mathcal{O}_2$).
Consider the scenario of injecting a single photon into mode H (denoted as $\ket{\widetilde{1,0}}$ in the main text by an abuse of notation), the true initial state is a mixed state of $\ket{0,0}$ and $\ket{1,0}$ with the probabilities of
\begin{equation}
	P^{\ket{\widetilde{1,0}}}_{\ket{0,0}}=1-\mathcal{O}_1, \quad P^{\ket{\widetilde{1,0}}}_{\ket{1,0}}=\mathcal{O}_1.
\end{equation}
If acting $U_g^{\text{PDC}}$ on the initial mixed state, the photon-number distribution of the H-polarized output mode is
\begin{equation}
	\begin{aligned}
        P_n^{\ket{\widetilde{1,0}}}&=
        P^{\ket{\widetilde{1,0}}}_{\ket{0,0}}*P_n^\mathrm{SPDC}+P^{\ket{\widetilde{1,0}}}_{\ket{1,0}}*P_n^{|1,0\rangle} \\
		&=\frac{(g-1)^{n-1}}{g^{n+1}}\Big[(1-\mathcal{O}_1)(g-1)+\mathcal{O}_1n\Big].
	\end{aligned}
 \label{eq:P10}
\end{equation}
With known $ P_n^{\ket{\widetilde{1,0}}}$, we can calculate the corresponding $m$-fold coincidence probabilities, which are parameterized by $g$ and $\mathcal{O}_1$. By substituting fitted $g$ acquired as mentioned in \ref{sec:g} into Eq.~\ref{eq:P10}, and then fitting the calculated $C_m$ with the measured $m$-fold coincidence probabilities, we get the overlap $\mathcal{O}_1$ of the incident single photon with the H-polarized mode of PDC.
\begin{figure}
\centering
\includegraphics[width=0.6\textwidth]{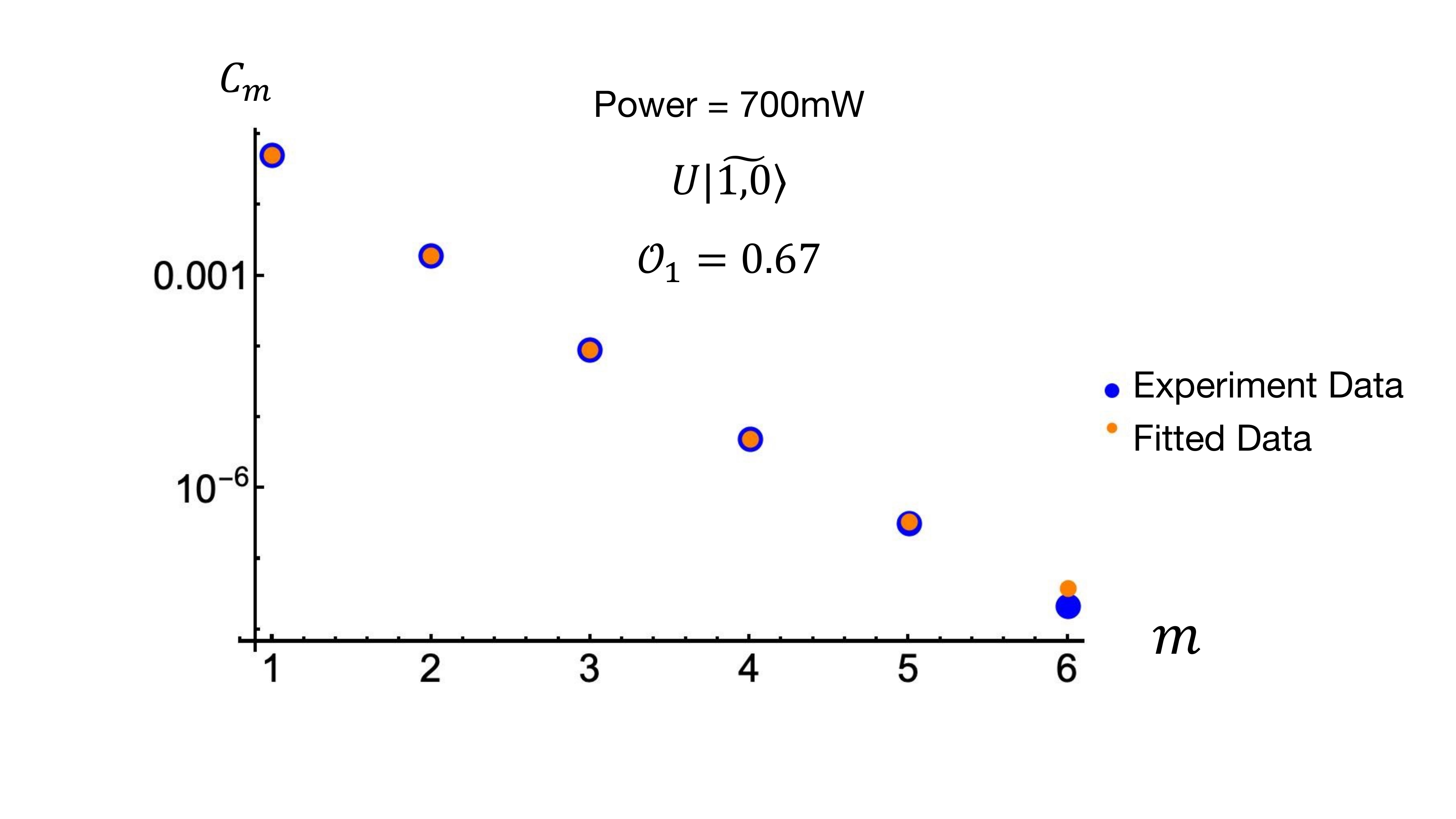}
\caption{Fitting the measured $m$-fold coincidence $C_m$ and calculated $C_m$ for the output state with photon number distribution shown in Eq.~\ref{eq:P10} to get the parameter $\mathcal{O}_1$.}
\label{fig:fit_10}
\end{figure}

\begin{figure}
\centering
\includegraphics[width=0.6\textwidth]{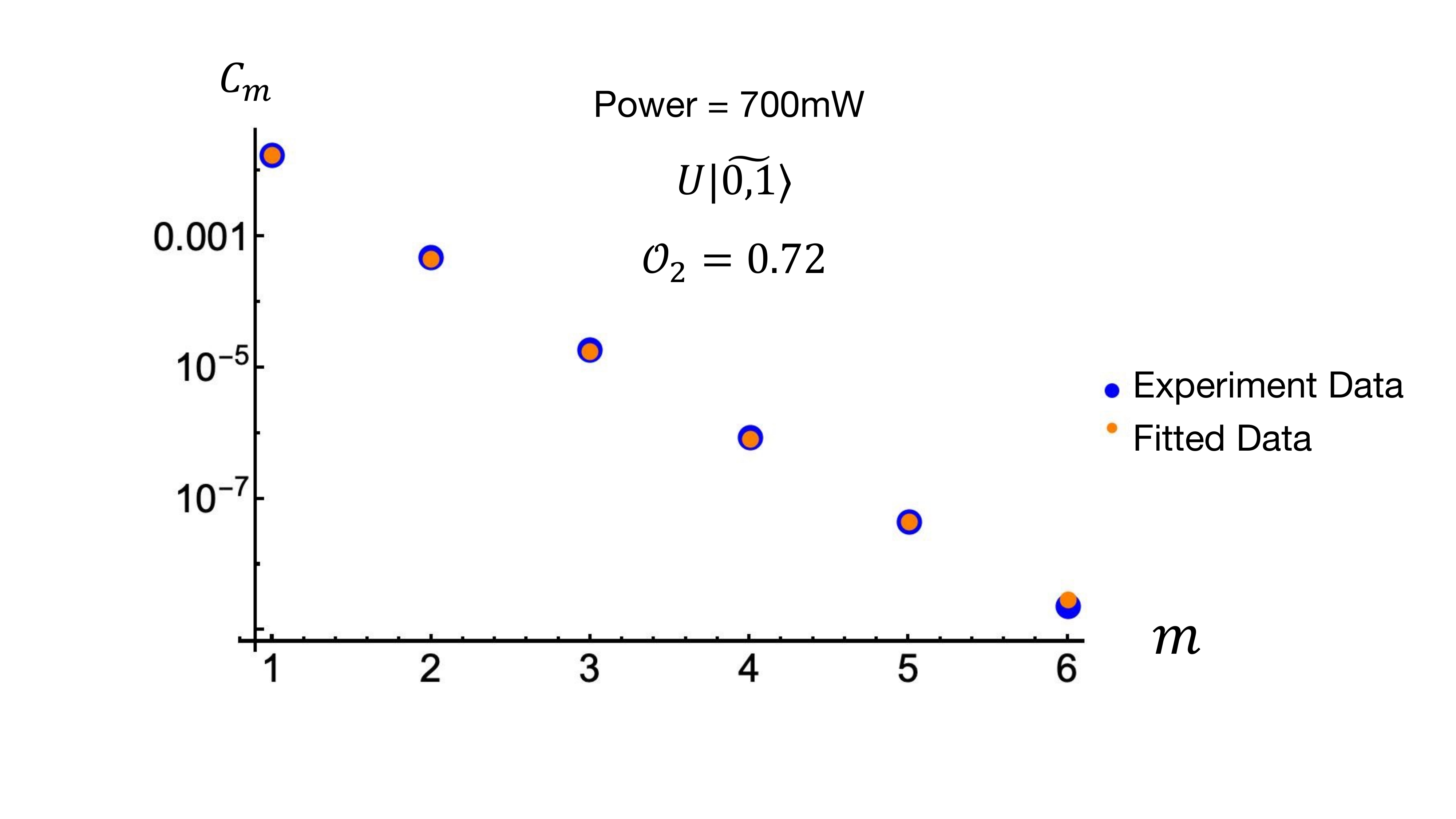}
\caption{Fitting the measured $m$-fold coincidence $C_m$ and the calculated $C_m$ for the output state with photon number distribution shown in Eq.~\ref{eq:p01} to get the parameter $\mathcal{O}_2$.}
\label{fig:fit_01}
\end{figure}

Likewise, if the overlap of the incident vertically polarized photon with the V-polarized mode in the PDC is $\mathcal{O}_2$, the initial state is a mixed state of $\ket{0,0}$ and $\ket{0,1}$ with the mixing fraction being
\begin{equation}
	P^{\ket{\widetilde{0,1}}}_{\ket{0,0}}=1-\mathcal{O}_2, \quad P^{\ket{\widetilde{0,1}}}_{\ket{0,1}}=\mathcal{O}_2,
\end{equation}
respectively.
The actual photon-number distribution of H-polarized output mode while injecting a photon into V-polarized mode is
\begin{equation}
	\begin{aligned}
		P_n^{\ket{\widetilde{0,1}}}&=
        P^{\ket{\widetilde{1,0}}}_{\ket{0,0}}*P_n^\mathrm{SPDC}+P^{\ket{\widetilde{1,0}}}_{\ket{0,1}}*P_n^{|0,1\rangle} \\
		&=\frac{(g-1)^{n}}{g^{n+2}}\Big[(1-\mathcal{O}_2)g+\mathcal{O}_2(n+1)\Big].
	\end{aligned}
 \label{eq:p01}
\end{equation}
The overlap of the V mode $\mathcal{O}_2$ also can be obtained from the fitting shown in Fig.~\ref{fig:fit_01}.

\subsection{PDC with initial state of $\ket{1,1}$}
For observing two-photon CJ nonlinear interference, the ideal initial state is $\ket{1,1}$, corresponding to impinge single photons in both modes at the same time. In realistic experiment, the input two single photons are always not perfectly overlapped with the PDC modes of PPKTP3. It is reasonable to assume that the matching of incident H-polarized photon with the H-polarized PDC mode of the PPKTP3 and matching of the incident V-polarized photon with the V-polarized mode of PPKTP3 are independent. Therefore, the actual incident state of ${\ket{\widetilde{1,1}}}$ is a mixed state of $\ket{0,0}$, $\ket{0,1}$, $\ket{1,0}$ and $\ket{1,1}$ with the probabilities respectively being 
\begin{equation}
	\begin{aligned}
		P^{\ket{\widetilde{1,1}}}_{\ket{0,0}}&=(1-\mathcal{O}_1)(1-\mathcal{O}_2), \\
		P^{\ket{\widetilde{1,1}}}_{\ket{0,1}}&=(1-\mathcal{O}_1)\mathcal{O}_2, \\
		P^{\ket{\widetilde{1,1}}}_{\ket{1,0}}&=\mathcal{O}_1(1-\mathcal{O}_2), \\
		P^{\ket{\widetilde{1,1}}}_{\ket{1,1}}&=\mathcal{O}_1\mathcal{O}_2. \\
	\end{aligned}
\end{equation}
After implementing $U_g^\text{PDC}$ on initial state ${\ket{\widetilde{1,1}}}$, the photon number distribution of the H-polarized output mode is
\begin{equation}
	\begin{aligned}
        P_n^{\ket{\widetilde{1,1}}}=&~P^{\ket{\widetilde{1,1}}}_{\ket{0,0}}*P_n^\mathrm{SPDC}+P^{\ket{\widetilde{1,1}}}_{\ket{0,1}}*P_n^{|0,1\rangle}+P^{\ket{\widetilde{1,1}}}_{\ket{1,0}}*P_n^{|1,0\rangle}+P^{\ket{\widetilde{1,1}}}_{\ket{1,1}}*P_n^{|1,1\rangle} \\
		=&~\frac{(g-1)^{n-1}}{g^{n+2}}\Big[(1-\mathcal{O}_1)(1-\mathcal{O}_2)(g-1)g+(1-\mathcal{O}_1)\mathcal{O}_2(g-1)(n+1) \\
		&~+\mathcal{O}_1(1-\mathcal{O}_2)gn+\mathcal{O}_1\mathcal{O}_2(n+1-g)^2\Big]. \\
	\end{aligned}
 \label{eq:p11tildle}
\end{equation}

\begin{figure}
\centering
\includegraphics[width=0.6\textwidth]{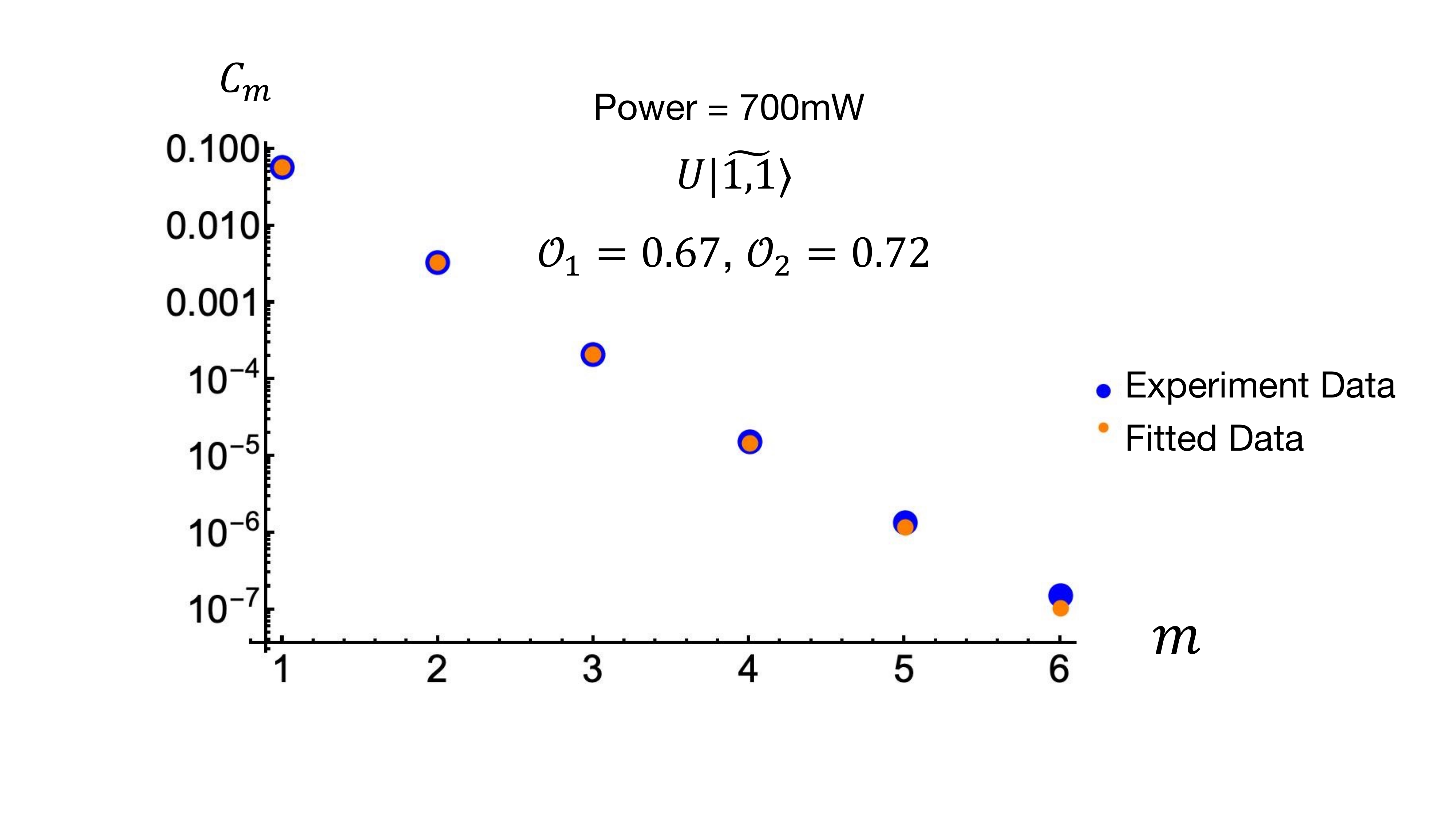}
\caption{Measured $m$-fold coincidence $C_m$ and calculated results from the fitted parameters for the output state when two single photons are initially injected.}
\label{fig:nofit_11}
\end{figure}

No additional parameters are needed here to describe the output state. Since the parameters $g$, $\mathcal{O}_1$ and $\mathcal{O}_2$ have obtained from auxiliary experiments as we discussed in \ref{sec:g} and \ref{sec:mathcal{O}}, the output photon number distribution of two-photon nonlinear interference experiment can be calculated by substituting the fitted parameters into Eq.~\ref{eq:p11tildle}. In Fig.~\ref{fig:nofit_11}, we show the calculated $C_m$ by substituting the fitted parameters into our model agrees well with the directly measured $C_m$ for a two-photon nonlinear experiment. In Fig.~4a of the main text, when $g=1.21$, the $P_1$ represented by the red dots are obtained via the $P_1^\text{5-detectors}$ and the red line is the $P_1$ calculated by substituting fitted parameter $g$, $\mathcal{O}_1$ and $\mathcal{O}_2$ into Eq.~\ref{eq:p11tildle}. These results verify that the model parameterized by gain $g$ and modes overlapping $\mathcal{O}_{1/2}$ we introduced in Section \ref{sec:deducingP1} can faithfully describe the two-photon nonlinear interference when two single photons are both input. In the case of $g\approx2$ where the approximation of $P_1$ by $P_1^\text{5-detectors}$ is invalid, we deduce $P_1$ and even $P_n$ (Fig.~\ref{fig:Pn_g=2}) with arbitrary $n$ by calculating the output state of our experiment model with fitted parameters.

Here we give the summary of getting important parameters from different measurement settings.
\begin{itemize}
	\item detection efficiency $\eta$ : calculated from counts of each single mode and coincidence counts of both H-polarized and V-polarized modes.
	\item nonlinear gain $g$ :  fitted from $m$-fold coincidence of the H-polarized output mode of SPDC when a specific pump power is chosen.
	\item overlap of the incident H polarized single photon with the H-polarized PDC mode of PPKTP3 $\mathcal{O}_1$ : fitted from $m$-fold coincidence of the H-polarized output mode of the PDC crystal with initial state of $|\widetilde{1,0}\rangle$
	\item overlap of the incident V polarized photon with the V-polarized PDC mode of PPKTP3 $\mathcal{O}_2$ : fitted from $m$-fold coincidence of the H-polarized output mode of the PDC with initial state of $|\widetilde{0,1}\rangle$
\end{itemize}

The experimental data for simultaneous inputting two photons are in good agreement with the calculated results obtained from parameterized model, which strongly proves the validity of the model.

\subsection{Detailed considerations}
In our experiment, the incident single photons are prepared by triggering the counterpart photon generated in SPDC process. However, the heralded states are not perfect single photon states, and they have some probabilities to be multi-photon states. Here we correct the heralded states so that the output states of the fitted model introduced above will be closer to the real output states in experiment.

Assuming the gain of SPDC process in PPKTP1 is gain $g_1$, the probability of outputting $n$ pairs of photons is
\begin{equation}
	p^{\mathrm{SPDC}1}_n=\frac{(g_1-1)^n}{g_1^{n+1}}.
\end{equation}
After photons in the V-polarized mode is triggered by detector Tig-1 with the trigger efficiency of the $\eta_{T1}$, the photon number distribution of the heralded H-polarized state that can interact with PPKTP3 is
\begin{equation}
	\begin{aligned}
		P_{n|\mathrm{Trig1}}&=\frac{P_{n \& \mathrm{Trig1}}}{P_\mathrm{Trig1}} \\
		&=\frac{\sum_{m=n}^\infty p^{\mathrm{SPDC} 1}_m\frac{m!}{n!(m-n)!}\mathcal{O}_1^n(1-\mathcal{O}_1)^{m-n}\Big[1-(1-\eta_{T1})^m\Big]}{\sum_{m=1}^\infty p^{\mathrm{SPDC} 1}_m\Big[1-(1-\eta_{T1})^m\Big]}, \\
	\end{aligned}
 \label{eq:pnT1}
\end{equation}
where $\mathcal{O}_1$ the overlap of the heralded H-polarized photon with the H-polarized PDC mode of PPKTP3. 
Similarly, we can calculate the photon number distribution of heralded V-polarized state from PPKTP2. If the gain of the SPDC in PPKTP2 is represented by $g_2$, the probability of outputting $n$ pairs of photons from PPKTP2 is
\begin{equation}
	p^{\mathrm{SPDC} 2}_n=\frac{(g_2-1)^n}{g_2^{n+1}}.
\end{equation}
After photons in the H-polarized mode is triggered by detector Tig-2 with the trigger efficiency of the $\eta_{T2}$, the photon number distribution of the heralded V-polarized state that can interact with PPKTP3 is
\begin{equation}
	\begin{aligned}
		P_{n|\mathrm{Trig2}}&=\frac{P_{n \& \mathrm{Trig2}}}{P_\mathrm{Trig2}} \\
		&=\frac{\sum_{m=n}^\infty p^{\mathrm{SPDC} 2}_m\frac{m!}{n!(m-n)!}\mathcal{O}_2^n(1-\mathcal{O}_2)^{m-n}\Big[1-(1-\eta_{T2})^m\Big]}{\sum_{m=1}^\infty p^{\mathrm{SPDC} 2}_m\Big[1-(1-\eta_{T2})^m\Big]}, \\
	\end{aligned}
 \label{eq:pnT2}
\end{equation}
where $\mathcal{O}_2$ the overlap of the heralded V-polarized photon with the V-polarized PDC mode of PPKTP3. 

Since parameters $\eta_{T1/2}$, $g_{1/2}$, $\mathcal{O}_{1/2}$ are all measurable in auxiliary experiments, we can correct the initial state that interacts with PPKTP3 by using Eq.~\ref{eq:pnT1} and \ref{eq:pnT2} to get a more accurate output state from PPKTP3. If the input state having $j$ photons in H-polarized and $k$ photons in V-polarized mode, the probability of outputting $n$ H-polarized photons from PPKTP3 is
\begin{equation}
	\begin{aligned}
		P_{n|jk}&=\Big|\langle n,m|U_g^\mathrm{PDC}|j,k\rangle\Big|^2 \\
		&=\frac{1}{g}\Big|\langle n,k|U_{1/g}^\mathrm{BS}|j,m\rangle\Big|^2 \\
		&=\frac{1}{g}\frac{k!}{n!j!(n+k-j)!}\frac{1}{g^{n-k}}\Big[\Big(\frac{\partial}{\partial\alpha}\Big)^n[\alpha^j(1+\alpha\beta)^{n+k-j}]\Big|_{\alpha=-\sqrt{\eta(1-\eta)},\beta=\sqrt{\frac{1-\eta}{\eta}}}\Big]^2 \\
		&=\frac{1}{g}\frac{k!}{n!j!(n+k-j)!}\frac{1}{g^{n-k}}\Big[\Big(\frac{\partial}{\partial\alpha}\Big)^n[\alpha^j(1+\alpha\beta)^{n+k-j}]\Big|_{\alpha=-\sqrt{g-1}/g,\beta=\sqrt{g-1}}\Big]^2, \\
	\end{aligned}
\end{equation}
which satisfying $n-m=j-k$~\cite{nakazato_photon_2016}. For some specific cases that initially only having either H-polarized photons or V-polarized photons, we have
\begin{equation}
	U_g^\mathrm{PDC}|j,0\rangle=\frac{1}{g^{(j+1)/2}}\sum_{n=0}^\infty\sqrt{\frac{(n+j)!}{n!j!}}\Big(\frac{g-1}{g}\Big)^{n/2}|n+j,n\rangle,
\end{equation}
\begin{equation}
	P_{n|j0}=\frac{n!}{j!(n-j)!}\frac{(g-1)^{n-j}}{g^{n+1}};
\end{equation}
or
\begin{equation}
	U_g^\mathrm{PDC}|0,k\rangle=\frac{1}{g^{(k+1)/2}}\sum_{n=0}^\infty\sqrt{\frac{(n+k)!}{n!k!}}\Big(\frac{g-1}{g}\Big)^{n/2}|n,n+k\rangle,
\end{equation}
\begin{equation}
	P_{n|0k}=\frac{(n+k)!}{n!k!}\frac{(g-1)^{n}}{g^{n+k+1}}.
\end{equation}

Therefore, the corrected initial state consists of the $\ket{j,k}$ with a probability of $P_{j|\mathrm{Trig1}} \times P_{k|\mathrm{Trig2}} $. Then the actual probability of outputting $n$ photons from the H-polarized mode of PPKTP3 reads
\begin{equation}
	\begin{aligned}
	P_n&=\sum_{j=0}^\infty\sum_{k=0}^\infty P_{n|jk} \times P_{j|\mathrm{Trig1}} \times P_{k|\mathrm{Trig2}}. \\
	\end{aligned}
\end{equation}

\section{Extended Data }
For resource saving, only H-polarized mode is detected in experiment in the main text. To ensure V-polarized mode actually behaves as same as the H-polarized mode, we measured both the H-polarized and V-polarized modes of the output state of PPKTP3 pumped by a pulsed laser of 700 mW. The approximately calculated $P_1$ by $P_1^\text{5-detectors}$ for H-polarized mode is $0.516\pm0.001$ and $0.383\pm0.008$ for the initial state of $\ket{\widetilde{1,0}}$ and  $\ket{\widetilde{1,1}}$, respectively. $P_1$ of $\ket{\widetilde{1,1}}$ decreases relative to the $P_1$ of $\ket{\widetilde{1,0}}$. It shows that when inputting extra V-polarized photon together with the H-polarized photon, the probability of outputting one H-polarized photon decreases, implying two-photon destructive interference occurs when single photons are injected in pairs. 

Similarly, the approximately calculated $P_1$ by $P_1^\text{5-detectors}$ for V-polarized mode is $0.489\pm0.001$ and $0.387\pm0.008$ for the initial state of $\ket{\widetilde{0,1}}$ and  $\ket{\widetilde{1,1}}$, respectively. It shows that when inputting extra H-polarized photon together with the V-polarized photon, the probability of outputting one V-polarized photon decreases, also implying two-photon destructive interference occurs when single photons are injected in pairs. 

\begin{figure}
\centering
\includegraphics[width=0.4\textwidth]{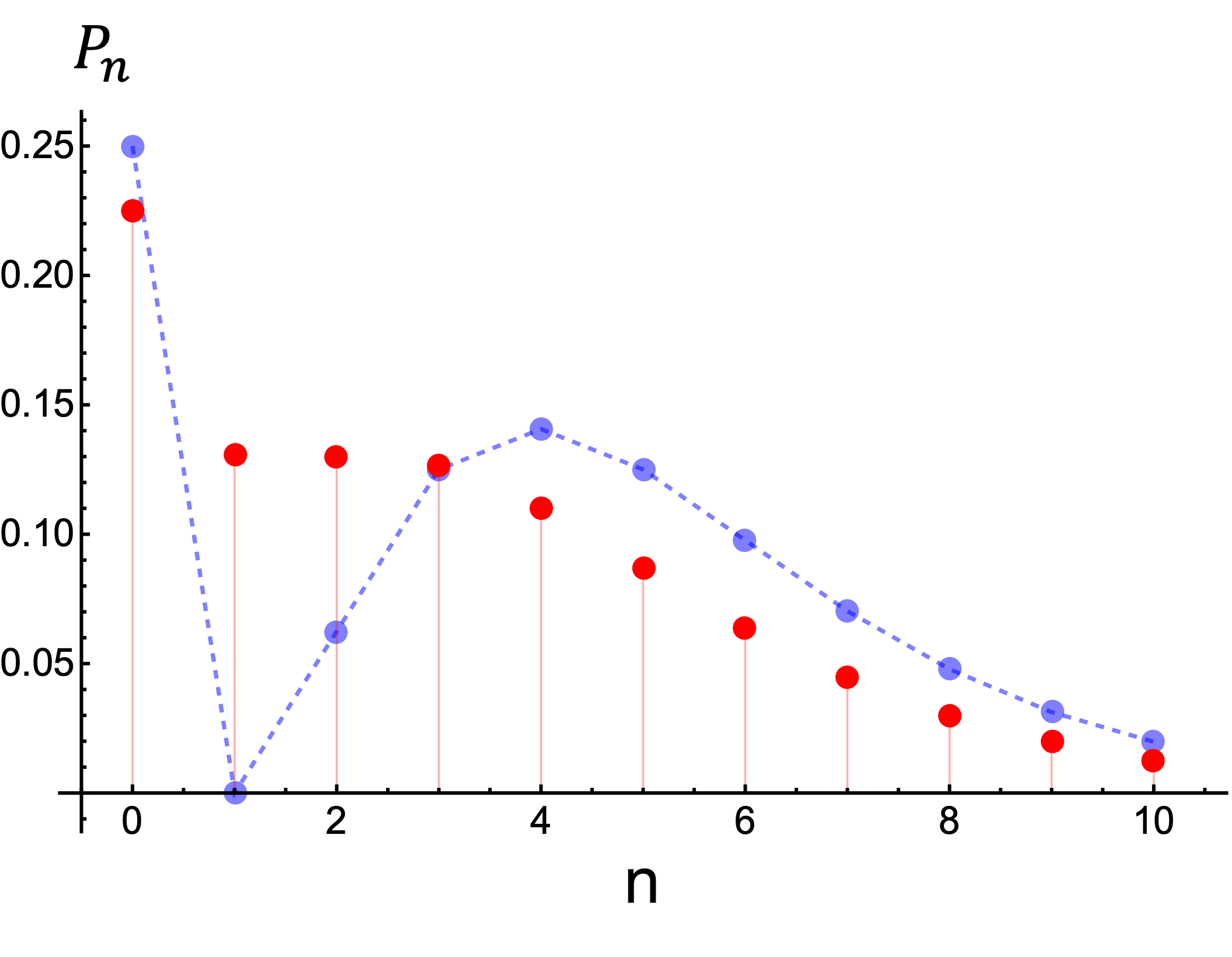}
\caption{Ideal (blue) and experimentally obtained (red) photon number distribution.
Experimental Data is deduced from the experimental model with fitted parameters $\mathcal{O}_1=0.65$, $\mathcal{O}_2=0.74$ and $g=2.03$. If all experimental conditions are perfect, i.e., $\mathcal{O}_1=1$, $\mathcal{O}_2=1$ and $g=2$, the probability $P_1$ vanishes ($P_1=0$).}
\label{fig:Pn_g=2}
\end{figure}

\begin{figure}
\centering
\includegraphics[width=0.95\textwidth]{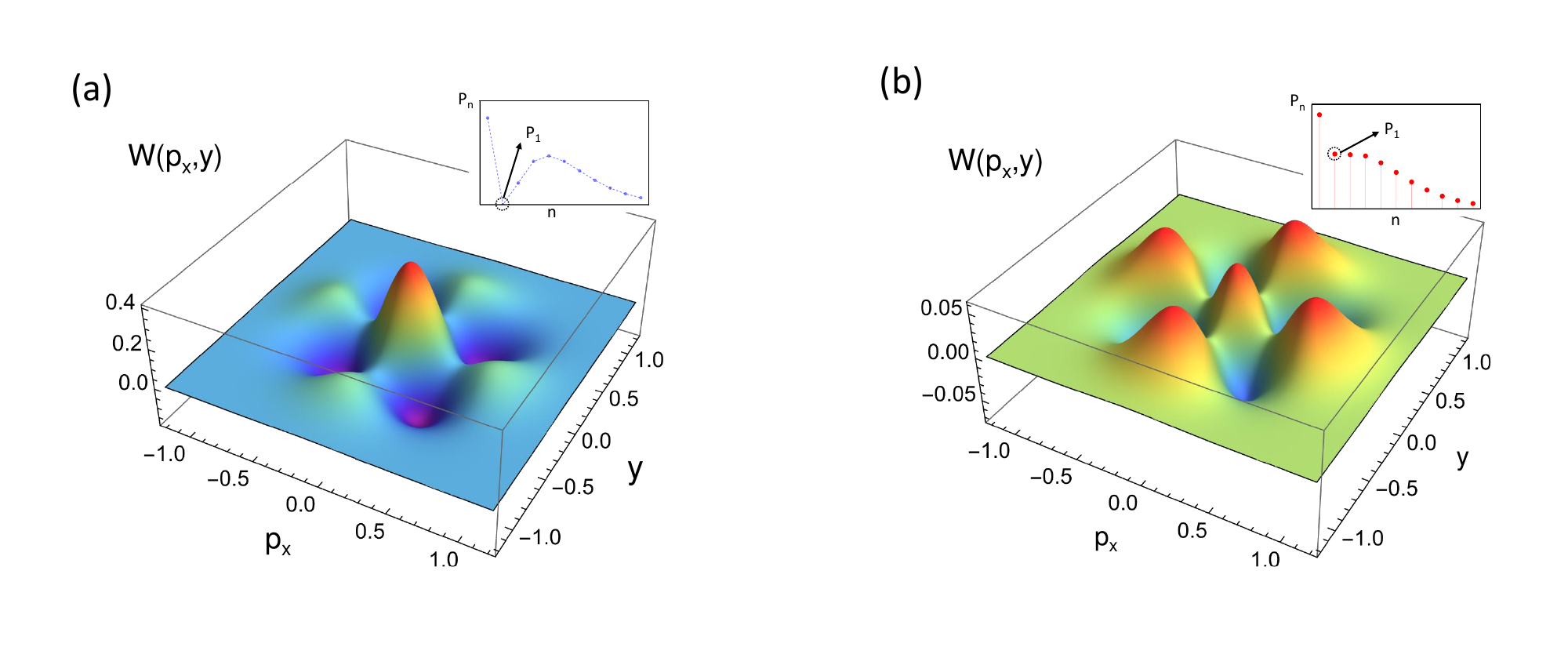}
\caption{Wigner function of the two-mode output state of CJ nonlinear interference when $g=2$. (a) Theoretically simulated Wigner function $W(p_x,y)$ with $x=0$ and $p_y=0$ of ideal output state of the CJ nonlinear interference. (b) Reconstructed Wigner function $W(p_x,y)$ with $x=0$ and $p_y=0$ of the output two-mode state in experiment reconstructed from our fitted experimental model. The insets of (a) and (b) show the corresponding photon-number distributions $P_n$, respectively.}
\label{fig:wigner_function}
\end{figure}

In Fig.~\ref{fig:Pn_g=2}, we show the deduced photon number distribution of the output state of two-photon CJ nonlinear interference experiment when $g=2.03$. In Fig.~\ref{fig:wigner_function}, we show the Wigner function $W(p_x,y)$ with $x=0$ and $p_y=0$ of the output two-mode state of two-photon CJ nonlinear interference.

\bibliography{suppreference}
\bibliographystyle{apsrev4-2}